\documentclass[twocolumn,superscriptaddress,floatfix,prx]{revtex4-1}
\usepackage{graphicx,amsfonts,amssymb,amsmath,hyperref,hypcap,enumerate}
\usepackage{color}
\usepackage{cleveref}

\newcommand{\qq}{\boldsymbol{q}}

\newcommand{\rr}{\boldsymbol{r}}

\newcommand{\kk}{\boldsymbol{k}}

\newcommand{\bb}{\boldsymbol{b}}

\begin{document}
\title{Phonon-induced giant linear-in-$T$ resistivity in magic angle twisted bilayer graphene: Ordinary strangeness and exotic superconductivity}

\author{Fengcheng Wu}
\affiliation{Condensed Matter Theory Center and Joint Quantum Institute, Department of Physics, University of Maryland, College Park, Maryland 20742, USA}

\author{Euyheon Hwang}
\affiliation{Condensed Matter Theory Center and Joint Quantum Institute, Department of Physics, University of Maryland, College Park, Maryland 20742, USA}
\affiliation{SKKU Advanced Institute of Nanotechnology and Department of Nano Engineering, Sungkyunkwan University, Suwon, 16419, Korea}

\author{Sankar Das Sarma}
\affiliation{Condensed Matter Theory Center and Joint Quantum Institute, Department of Physics, University of Maryland, College Park, Maryland 20742, USA}

\date{\today}

\begin{abstract}
		We study the effect of electron-acoustic phonon interactions in twisted bilayer graphene on resistivity in the high-temperature transport and superconductivity in the low-temperature phase diagram. We theoretically show that twisted bilayer graphene should have an enhanced and strongly twist-angle dependent linear-in-temperature resistivity in the metallic regime with the resistivity magnitude increasing as the twist angle approaches the magic angle.  The slope of the resistivity versus temperature could approach one hundred ohms per kelvin with a strong angle dependence, but with a rather weak dependence on the carrier density. This higher-temperature density-independent  linear-in-$T$ resistivity crosses over to a $T^4$ dependence at a low density-dependent characteristic temperature, becoming unimportant at low temperatures. This  angle-tuned resistivity enhancement arises from the huge increase in the effective electron-acoustic phonon coupling in the system due to the suppression of graphene Fermi velocity induced by the flatband condition in the moir\'e superlattice system.  Our calculated temperature dependence is reminiscent of the so-called `strange metal' transport behavior except that it is arising from the ordinary electron-phonon coupling in a rather unusual parameter space due to the generic moir\'e flatband structure of twisted bilayer graphene. 
		We also show that the same enhanced electron-acoustic phonon coupling also mediates  effective  attractive interactions in $s$, $p$, $d$ and $f$ pairing channels with a theoretical superconducting transition temperature on the order of $\sim$5 K near magic angle. The fact that ordinary acoustic phonons can produce exotic non-$s$-wave superconducting pairing arises from the unusual symmetries of the system.
\end{abstract}

\maketitle

\section{introduction}
The remarkable recent discovery\cite{Cao2018Super,Cao2018Magnetic} of the existence of correlated insulating phases and superconductivity in twisted bilayer graphene (TBG) has ushered in a new era in solid state physics where the twist angle $\theta$ between two two-dimensional (2D) layers of graphene serves as the tuning parameter, similar to carrier density ($n$) or magnetic field ($B$), in controlling the electronic properties of a system.  The initial dramatic discovery has already been reproduced and expanded in follow-up publications from other groups \cite{Dean2018Tuning}. A huge number of theoretical papers have followed the experimental discovery propounding many different aspects of TBG \cite{Balents2018,roy2018unconventional, Senthil2018,Koshino2018, Kang2018,  Dodaro12018, Padhi2018, Guo2018, huang2018AF, Liu2018chiral,Fidrysiak2018, Heikkila2018, Kennes2018strong, Isobe2018, You2018, rademaker2018charge,PALee2018, Wu2018phonon, guinea2018electrostatic,Carr2018, Thomson2018, gonzalez2018kohn, Lin2018, Lian2018twisted, Ahn2018failure, sherkunov2018novel,Venderbos2018, choi2018electron}, and the subject is in a flux with many more results soon to follow.  In particular, a consensus, by no means universal, is developing that the ground states of TBG near the magic angle are `strongly correlated' and `Mott-like' in the sense that many-body interaction effects are driving the 2D metallic Fermi liquid ground state into a gapped insulator with a superconducting instability developing as one dopes slightly away from these insulating states, similar perhaps to what happens in cuprate high-temperature superconductors.  If so, then studying the electronic properties of TBG could shed light not only on the properties of strongly interacting graphene, but also perhaps of strongly interacting cuprates (and other strongly correlated 2D materials) where strong correlation produces interesting non-perturbative physics. Understanding properties of TBG therefore transcends just the physics of graphene and takes us possibly into the regime of strongly correlated materials where theoretical progress has been challenging.

The current work takes a  different perspective and asks about the nature of TBG transport properties at a higher temperature ($T$) and away from the moir\'e miniband edge, where the system behaves as an effective metal with the resistivity $\rho(T)$ increasing with increasing $T$ \cite{Cao2018Super,Cao2018Magnetic,Dean2018Tuning}.  This metallic behavior \cite{Cao2018Super,Cao2018Magnetic,Dean2018Tuning} happens for $T>T_m$, where $T_m (n, \theta)$ on the order of few kelvins is a density and angle dependent crossover temperature with the subscript $m$ denoting `metal', i.e., $d\rho/dT>0$ for $T>T_m$.  One of our primary goal is to understand the behavior of $\rho(T, n, \theta)$ in this $T>T_m$ effective metallic regime. We emphasize that $T_m$ is above the temperature scales of the superconducting or the correlated insulating ground states of the system.  The behavior of resistivity in this TBG effective metallic phase is of fundamental importance because the corresponding situation in cuprates (the metallic resistivity above the superconducting dome) has been dubbed a `strange metal' which is thought to be a hallmark of a non-Fermi-liquid (NFL) with very high resistivity and a linear-in-$T$ resistivity persisting to rather low temperatures.  The properties and the origin of strange metals are theoretically not understood, but are thought to arise from the strongly interacting nature of the corresponding strongly correlated ground state beyond any Fermi-liquid (FL) paradigm.  It seems that the qualitative behavior of TBG $\rho(T)$ is similar to that of the cuprates with a very high (thousands of times larger than that of regular monolayer graphene) and linear-in-$T$ resistivity reminiscent of a putative NFL strange metal \cite{MIT2018_rho, Columbia2018_rho} .  If indeed the TBG is a strange metal for $T>T_m$, this would be a tremendous advance in our understanding of strange metals since the transport properties of the parent material (i.e. graphene) are  well-understood both qualitatively and quantitatively \cite{CastroNeto2009,Peres2010, DasSarma2011}. There is also a theoretical possibility that the similarity between the phase diagrams of TBG and cuprates (i.e. nearby occurrences of insulating and superconducting phases as a function of doping as well as the presence of a higher-temperature linear-in-$T$ resistivity above the superconducting region) could be accidental, and one ultimate goal of theory should be to clarify possible fundamental connections (or not) between these two very different classes of systems.  Our theoretical work is entirely on TBG, and we mention cuprates because the similarity between TBG and cuprate properties was pointed out in Refs.~\cite{Cao2018Super}.

In the current work we develop a quantitative, albeit approximate, theory for $\rho (T, n, \theta)$ in TBG by considering electron-phonon scattering within the context of continuum TBG moir\'e band structure.  We neglect all effects of disorder and electron-electron interaction in this zeroth order theory assuming that the system is extremely clean (‘no disorder’) and that electron interaction does not directly affect transport at the finite doping samples of interest in the TBG metallic regime (i.e. we stay slightly away from the Dirac point since the TBG strange metal behavior manifests itself only at finite carrier densities away from the charge neutral Dirac point).  We also neglect all phonon modes other than the low energy in-plane acoustic phonon modes in our consideration.  Our reasons for neglecting other phonon modes are the following: (1) the in-plane acoustic phonons by themselves provide a quantitatively accurate description for the temperature dependent resistivity in regular monolayer graphene \cite{Hwang2008, Kim2010}; (2) by keeping just one kind of phonons, which are undoubtedly present in the system and are operational at low energies by virtue of their long wavelength acoustic dispersion (i.e. no gap as for optical phonon modes), we provide a clean parameter-free theoretical description, which is easy to compare with experiment and therefore can be improved in the future by including more complicated phonon modes, if necessary.

We show that the incorporation of the TBG band structure into the electron-phonon scattering theory produces a $\rho (T, n, \theta)$, which has a linear-in-$T$ behavior down to $T \sim 10 $ K (depending on density) and has a theoretically calculated resistivity value, using monolayer graphene parameters for electron-phonon coupling and sound velocity, on the order of 500 ohms  at $T$  $ \sim 5 - 10$ K. (This is roughly 1000 times larger than the corresponding monolayer graphene resistivity induced by exactly the same electron-acoustic phonon scattering at these temperatures.)  Our calculated $d\rho/dT$  reaches order of 100 ohms/K for  $\theta \sim 1.1^{\circ}$. In our model calculation, the magic angle $\theta_{\text{magic}}$ is about $1.025^{\circ}$ where the TBG Dirac velocity vanishes \cite{Bistritzer2011}. (The precise theoretical value of the magic angle depends on the model Hamiltonian and parameter values. Experiments \cite{Cao2018Magnetic, Cao2018Super, Dean2018Tuning} suggest that the magic angle is between $1^{\circ}$ and $1.2^{\circ}$.)   These findings appear to be in qualitative and semiquantitative agreement with experimental results \cite{Cao2018Super, Cao2018Magnetic, Dean2018Tuning, MIT2018_rho, Columbia2018_rho}. Our theoretically obtained $T_L$ ($>T_m$), above which the linearity in $\rho(T)$ develops, is a few kelvins depending on the doping density.   Thus, we argue that there is a reasonable case to be made that the main contribution to the strange metal TBG behavior arises from electron-phonon interaction within the FL paradigm of doped graphene although we certainly cannot rule out additional effects arising from strong correlations since electron-electron interactions (as well as disorder and additional phonon modes) are neglected a priori in our theory. Our calculated resistivity should be considered a lower bound to the measured values, and indeed, this appears to be the case.\cite{MIT2018_rho, Columbia2018_rho}

A second, and equally important, part of our work is the investigation of TBG superconductivity arising from electron-acoustic phonon interaction in the flatband moir\'e superlattice situation.
We  use the same electron-acoustic phonon coupling to study the superconducting instability at low temperatures, motivated by the possibility that the high-temperature phonon induced giant linear-in-$T$ resistivity and the low-temperature superconductivity could arise from the same electron-phonon interaction. This is well-known in the context of normal metals (e.g. Al, Pb) where the electron-acoustic phonon couplings extracted from the high-temperature resistivity agree well (usually within 10\%) with the ones extracted from tunneling measurements in the superconducting state.\cite{PAllen}
In graphene, however, the presence of the pseudospin modifies the situation in a subtle manner, and the dimensionless electron-phonon couplings obtained from the high-temperature resistivity and the superconducting $T_c$ differ by a factor of $2-4$ as we will show in the current work.
Using a microscopic theory for superconductivity, we show that acoustic phonons generate an effective attractive interaction not only in the conventional $s$-wave channel, but also in the unconventional $p$-, $d$- and $f$-wave channels; the attraction in flatbands near magic angle is strong enough to produce a superconducting transition temperature $T_c$ on the order of $\sim$ 5 K. Within this model, spin singlet $s$-wave and spin triplet $f$-wave pairings have degenerate $T_c$ because each valley has its own spin rotational symmetry; for the same reason, $p$- and $d$-wave pairings are also degenerate. Our study presents a novel example where conventional phonon fluctuations can lead to exotic unconventional pairings, including spin triplet channels. More experimental work will be necessary to determine the pairing symmetry of the reported TBG superconducting state, but our work points to the possibility of exotic superconductivity from an ordinary paring  mechanism just as our theory of resistivity points to a strange metallic behavior arising from an ordinary scattering mechanism.   We emphasize that both effects studied in this paper arise from moir\'e superlattice induced strong enhancement of the effective electron-phonon TBG coupling through the mechanism of the flatband induced strongly twist angle dependent suppression of the TBG Fermi velocity.  Thus, both physical effects (i.e. strange metallicity and exotic superconductivity) show very strong twist angle dependence.
We believe that our work clearly establishes the qualitative and quantitative importance of electron-acoustic phonon interaction being of great potential importance in the physics of TBG.

This paper is organized as follows. We present the theory of phonon-induced giant linear-in-$T$ resistivity (i.e. strange metallicity) in Sec.~\ref{sec:resistivity} and the theory of acoustic-phonon-mediated exotic superconductivity in Sec.~\ref{sec:SC}. 
The same electron-acoustic phonon interactions, as specified in Sec.~\ref{sec:SC}, are used in both the resistivity and superconductivity theory.
We include both the layer symmetric and  antisymmetric acoustic phonon modes, because both are gapless in TBG, as explained in  Sec.~\ref{sec:SC}.
In the theory of resistivity, we use an effective Dirac Hamiltonian  with a renormalized velocity for electron energy spectrum, in order to obtain analytical results.
The Dirac dispersion is an approximation that is valid for Fermi energy near the charge neutrality point (Dirac point). Therefore, we limit our study of resistivity to low carrier density (below $10^{12}$ cm$^{-2}$). 
However, in the theory of superconductivity, we use the actual moir\'e band structure throughout the moir\'e Brillouin zone. All details of the moir\'e band structure, including Van Hove singularities, are faithfully incorporated into the superconductivity theory. This allows us to calculate the superconducting critical temperature $T_c$ as a function of filling factor for the full doping range of the flatbands. 
Finally, a brief conclusion is given in Sec.~\ref{sec:Discussion}. Most of the technical details of the theoretical calculations are given in  Appendices \ref{AppA}, \ref{AppB} and \ref{AppC},  so that the main part of the paper emphasizes the experimental consequences and the physical mechanisms underlying our theory without the clutter of too many equations.

\section{Phonon-induced giant linear-in-$T$ resistivity}
\label{sec:resistivity}

\begin{figure}[t]
	\includegraphics[width=1.0\columnwidth]{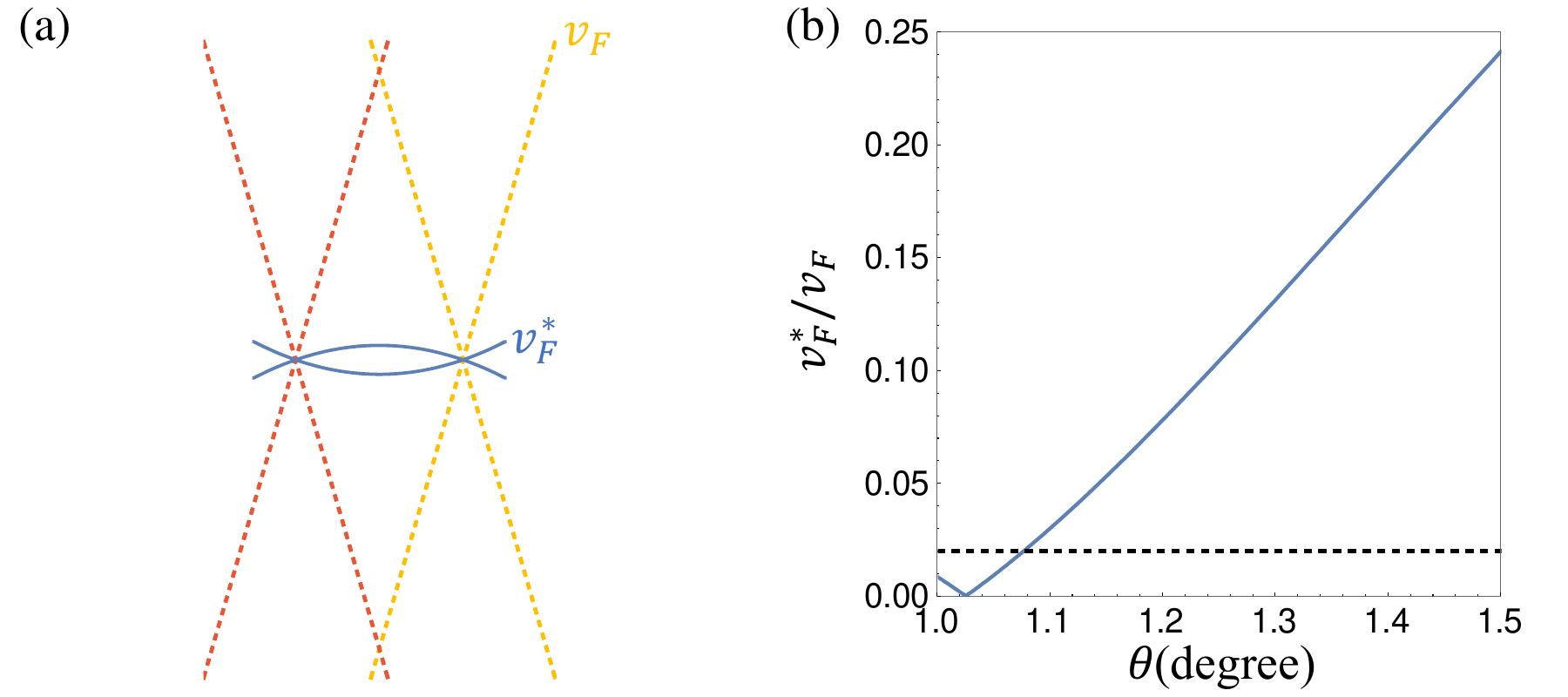}
	\caption{(a)Schematic illustration of the moir\'e band structure (blue solid lines) in TBG. The  dashed lines represent the pristine Dirac cones associated respectively with top and bottom graphene layers in the same valley. The interlayer hybridization reduces the velocity at Dirac points from the monolayer value $v_F$ to the renormalized value $v_F^*$. (b) $v_F^*/v_F$ as a function of $\theta$ from a full moir\'e band structure calculation as described in Appendix~\ref{AppA}. The horizontal dashed line indicates the phonon velocity $v_{ph}/v_F$, which is assumed to remain the same when $\theta$ changes in our theory.}
	\label{Fig:vF}
\end{figure} 

We start by a brief review of  the standard theory for acoustic phonon scattering induced $\rho (T,n)$ in monolayer graphene (MLG) developed by two of the authors \cite{Hwang2008, Min2011} and well-verified experimentally\cite{Kim2010}. This theory shows that the ‘high-temperature’ resistivity of MLG goes as:
\begin{equation}
\rho_{\text{MLG}}= \frac{\pi D^2}{ g_s g_v e^2 \hbar \rho_m  v_F^2 v_{ph}^2}k_B T
\label{rho_graphene}
\end{equation}
where $D$, $v_F$, $v_{ph}$, $\rho_m$ are respectively the graphene deformation potential from electron-phonon coupling, Dirac velocity, phonon (or sound) velocity, and atomic mass density.  
$g_s$ and $g_v$ respectively account for the spin and valley degeneracy, and both are 2 in graphene.
Equation~(\ref{rho_graphene}) applies both below and above the Fermi temperature $T_F$ of graphene as long as the phonons are in the high-temperature thermal equipartition regime, i.e., for $T>T_{BG}=2\hbar v_{ph} k_F /k_B $, where $T_{BG}$ is the so-called Bloch-Gr\"uneisen temperature  defining the energy of acoustic phonons having a wavenumber equal to twice of the Fermi wavenumber $ k_F$.  In reality, Eq.~(\ref{rho_graphene}) already applies for $T>T_{BG}/4$.\cite{Hwang2008, Min2011}  For low temperatures, $\rho(T)$ arising from electron-phonon scattering decreases strongly with $\rho \sim T^4$ for $T \ll T_{BG}$ \cite{Hwang2008,Kim2010, Min2011}.  This is the  Bloch-Gr\"uneisen regime where electron-phonon scattering is highly inefficient in causing resistive scattering as backscattering by phonons is strongly suppressed due to the quantum reduction in the phonon thermal occupancy.  The Bloch-Gr\"uneisen regime is not particularly relevant for the current work as we are interested in understanding the higher temperature linear-in-$T$ transport for $T>T_{BG}/4$ where electron-phonon scattering should dominate. Other resistive scattering mechanisms (e.g. disorder, electron-electron interactions) are likely to dominate the low-temperature resistivity where $\rho(T)$ is no longer linear in temperature.

It turns out, as shown in this work, that Eq.~(\ref{rho_graphene}) formally applies to TBG as the leading-order high-temperature resistivity formula as long as $v_F$ (and other parameters such as $D$) are redefined taking into account the moir\'e superlattice structure of the twisted system (Appendix~\ref{AppA}).
The main effect in TBG is a drastic reduction in the Fermi velocity near the magic angle \cite{Bistritzer2011}.  In particular, in an ideal situation (not realized experimentally so far), the renormalized Dirac velocity $v_F^{*}$ in TBG approaches zero at the magic angle, implying that $\rho (T)$ in Eq.~(\ref{rho_graphene}) would approach infinity as $1/v_F^{*2}$.  However, in the Dirac fermion model with linear dispersion, $v_F^*>v_{ph}$ is necessarily required to satisfy energy-momentum conservation in the intra-band electron-phonon scattering processes, as discussed in detail in Appendix~\ref{AppB}.  In practice, we find that Eq.~(\ref{rho_graphene}) provides a reasonable estimation when $v_F^{*} > 1.5 v_{ph}$  [see  Appendix~\ref{AppB}]. Therefore, here we focus on twist angles larger than 1.1$^\circ$, at which $v_F^{*} \approx 1.5 v_{ph} \approx 0.03 v_F $. Putting in the actual values of graphene parameters in Eq. (\ref{rho_graphene}), one finds that in MLG $d\rho/dT \sim$ 0.1 ohm/K for $T>T_{BG}$\cite{Hwang2008}, which also agrees with experiment\cite{Kim2010}, and therefore, in TBG the value of $d\rho/dT$ could be of order 100 ohms/K when $v_F^{*}  = 0.03 v_F$ is used.  If phonon velocity $v_{ph}$ in TBG is also renormalized, for example by a reduction of a factor of 2, $d\rho/dT$ values exceeding 1000 ohms/K are possible in TBG. A recent first-principles study \cite{choi2018electron} shows that the phonon spectrum of twisted bilayer graphene is insensitive to the moir\'e structure even at small twist angles ($\sim 1^{\circ}$).  Therefore, We do not consider any possible modification of the phonon properties by the moir\'e superlattice.


\begin{figure}[t]
	\includegraphics[width=1.0\columnwidth]{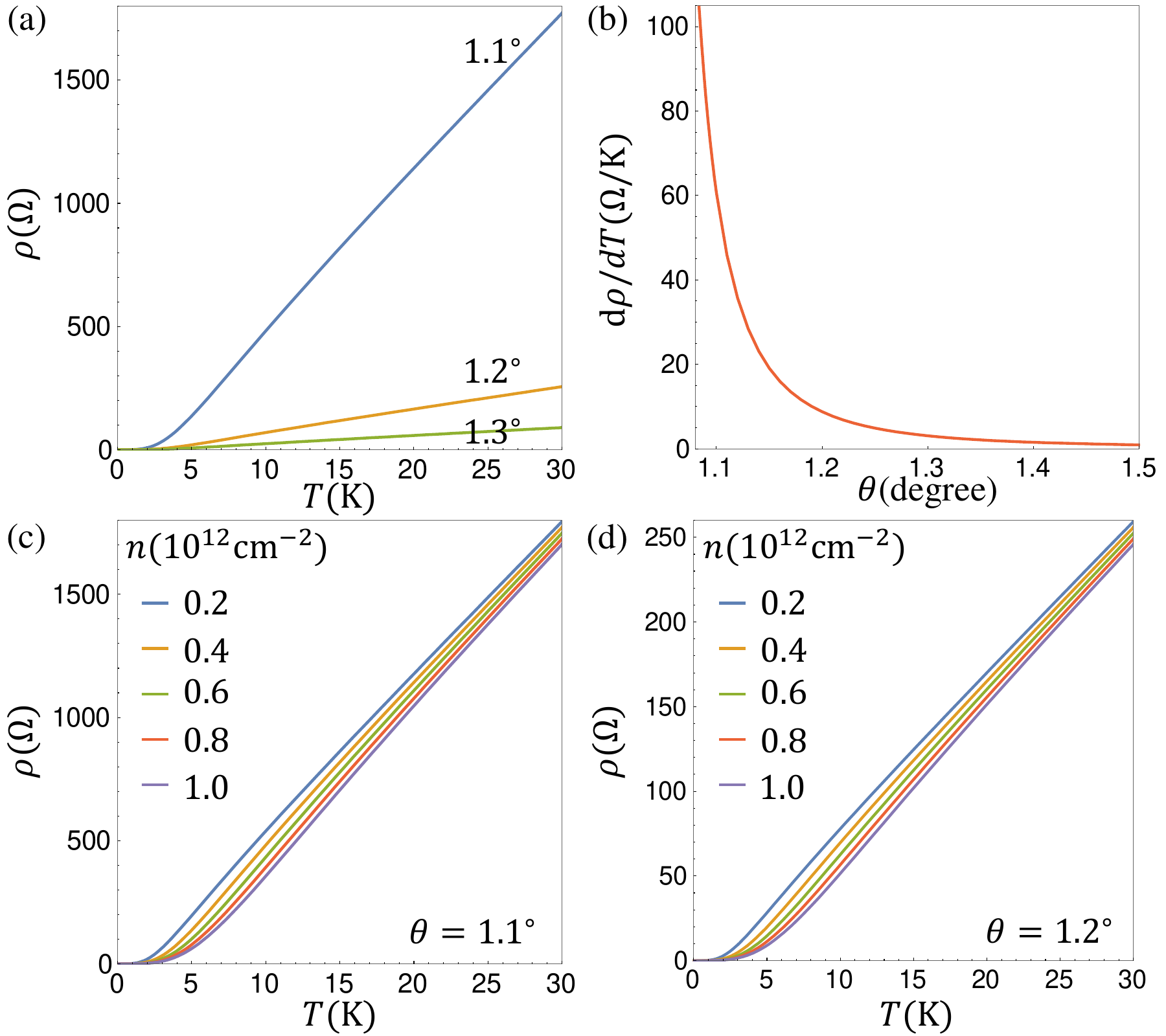}
	\caption{(a)$\rho(T)$ for three different twist angles. Electron density $n$ is $4\times 10^{11}$cm$^{-2}$  (b) $d\rho/dT$ as a function of $\theta$.  High temperatures $T>T_{BG}/4$ is assumed, so $d\rho/dT$ is independent of density based on Eq.~(\ref{rho_TBG}). (c) and (d) $\rho(T)$  for different electron density $n$. $\theta$ is $1.1^{\circ}$ in (c) and $1.2^{\circ}$ in (d). Monolayer graphene phonon parameters (given in the text) are used.}
	\label{Fig:rho}
\end{figure}

\begin{figure}[t]
	\includegraphics[width=1.0\columnwidth]{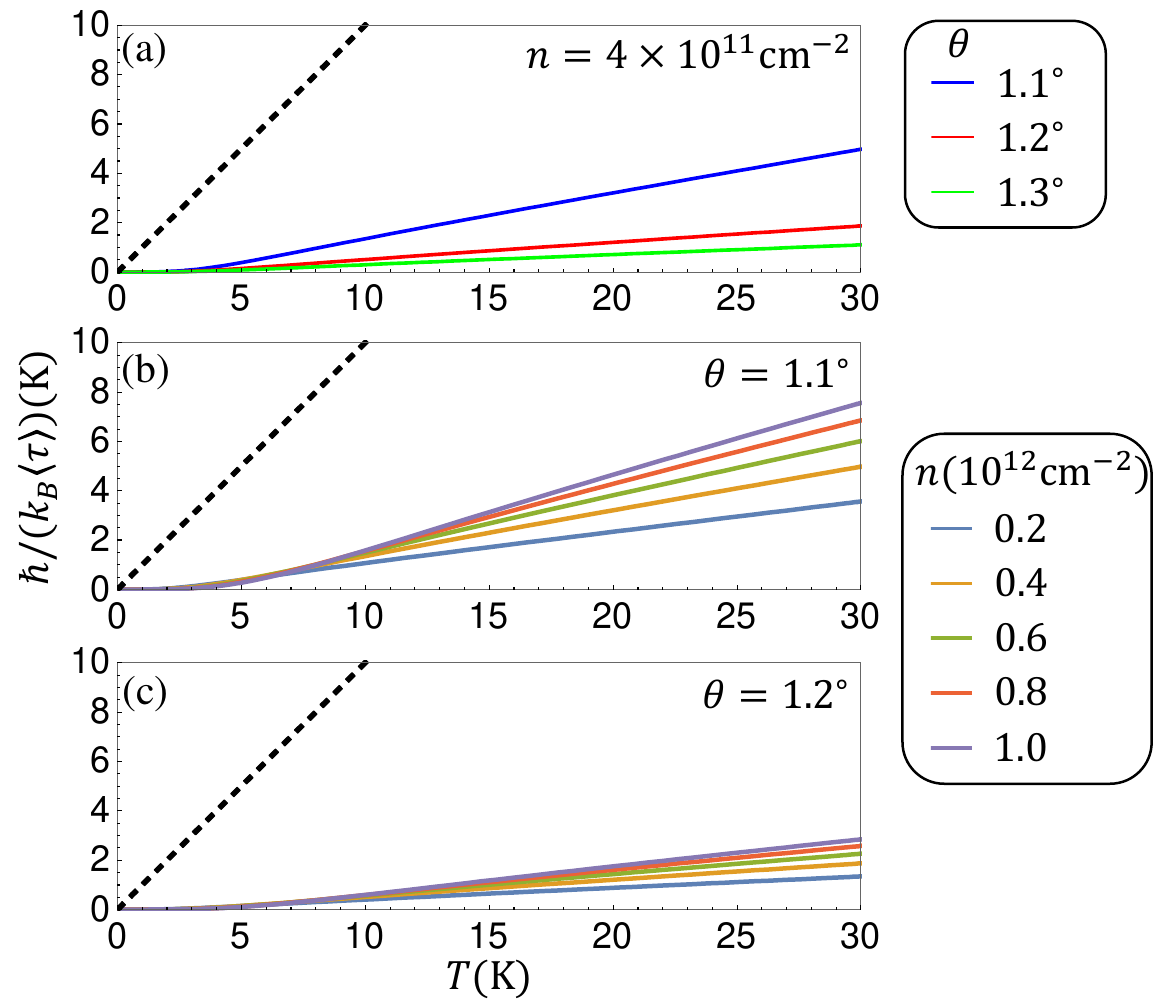}
	\caption{Scattering rate $\hbar/(k_B \langle \tau \rangle)$ as a function of temperature $T$. Different solid lines are for different twist angles in (a) but for different density $n$ in (b) and (c). In (a), $n$ is $4\times 10^{11}$cm$^{-2}$. $\theta$ is $1.1^{\circ}$ in (b) and $1.2^{\circ}$ in (c). The black dashed line in each plot represents the  Planckian scattering rate $\hbar/(k_B \langle \tau \rangle)=T$. Monolayer graphene phonon parameters (given in the text) are used.}
	\label{Fig:tau}
\end{figure}

In addition to the very large resistivity another characteristic of strange metals is the linear-in-$T$ resistivity persisting to rather low temperature.  This is guaranteed in our theory by the fact that $T_{BG}=2 \hbar v_{ph} k_F /k_B$ could be rather small at low carrier densities. Note that $T_{BG}$ is independent of graphene velocity.  Putting in $v_{ph}=2 \times 10^6$ cm/s  and $k_F=\sqrt{\pi n}$ appropriate for graphene, we find the Bloch-Gr\"uneisen temperature to be $T_{BG}=20 \sqrt{N}$ K where $N$ is the doping density (with respect to the charge neutrality point) measured in units of $10^{11}$ cm$^{-2}$.  Since the linearity-in-$T$ persists to roughly $T_L \sim T_{BG}/4$, we anticipate $\rho (T)$ to remain linear down to  $5 \sim 10$ K for $n=10^{11} \sim 10^{12}$ cm$^{-2}$. (We note that $T_m$ and $T_L>T_m$ in our notations are characteristic temperature scales above which $d\rho/dT>0$ and $d\rho/dT \sim$ constant, respectively.)  Thus, the resistivity of TBG is both extremely large and approximately linear down to low temperatures ($T_L$) in the metallic regime, with both effects arising from the gigantic enhancement in the effective electron-phonon coupling of the system due to the flatband induced suppression of the graphene Fermi velocity in the moir\'e band structure. The strange resistivity here is thus quite an `ordinary strangeness'.

We have carried out a calculation of the TBG metallic resistivity arising from electron-phonon interaction taking into account the TBG moir\'e band structure, explicitly verifying that Eq.~(\ref{rho_graphene}) applies to TBG as long as $v_F$ and other parameters are appropriately modified incorporating flatband moire superlattice effects.  The theory, described in Appendices \ref{AppA} and \ref{AppB}, includes both the actual Fermi velocity suppression and the modification of the electron-phonon matrix element due to the modified moir\'e  band structure as a function of the twist angle $\theta$, and the resistivity is then given by:
\begin{equation}
\begin{aligned}
\rho &= \frac{ 32 F(\theta) D^2 k_F}{ g_s g_v g_l e^2  \rho_m  v_F^{*2} v_{ph}} I(\frac{T}{T_{BG}}),\\
I(z)& = \frac{1}{z} \int_0^1 dx x^4 \sqrt{1-x^2} \frac{e^{x/z}}{(e^{x/z}-1)^2},
\end{aligned}
\label{rho_TBG}
\end{equation}
where $g_s$, $g_v$ and $g_l$ respectively account for the spin, valley and layer degeneracy and all are 2 in TBG. $g_l$ is  an extra factor of 2 compared with MLG arising from the two layers in TBG. Correspondingly $k_F$ is $\sqrt{\pi n/2}$, and $T_{BG}$ and $T_L$ in TBG are reduced by a factor of $\sqrt{2}$ with respect to MLG given the same density $n$. $v_F^*$ is the reduced Dirac velocity in TBG, which is obtained from the full moir\'e band structure calculation. The form factor $F(\theta)$ accounts for the change in electron-phonon matrix element of TBG, and is calculated  using a perturbation theory as shown in Appendix~\ref{AppA}. In general, we find $1>F>0.5$, and thus, the matrix element effect is not quantitatively large compared with the huge Fermi velocity suppression effect. Equation~(\ref{rho_TBG}) applies to low carrier density ($n\sim 10^{12}$ cm$^{-2}$ or below) so that continuum Dirac Hamiltonian approximation is still valid for TBG, and to velocity $v_F^*$ greater than $1.5 v_{ph}$ (equivalently, $\theta>1.1^{\circ}$) where quasielastic approximation works, as shown in Appendix~\ref{AppB}.  For $v_F^*< 1.5 v_{ph}$, Equation (\ref{rho_TBG}) still captures the qualitative behavior of $\rho(T)$, although it overestimates the value of $\rho$. A transport theory that is applicable for any ratio of $v_F^*/v_{ph}$ is given in Appendix~\ref{AppB}.

For large $\theta$, we reproduce the known results of Refs. \cite{Hwang2008, Min2011} for ordinary MLG, but for $\theta$ close to $1.1^{\circ}$ the resistivity results deviate appreciably from those in MLG.  We show our calculated results for $ \rho (T, n, \theta)$ in Figs. \ref{Fig:vF}-\ref{Fig:tau}. In Fig.~\ref{Fig:vF} the schematic TBG moir\'e band structure is shown.  Our interest is in a temperature range ($<$50 K) much smaller than the band gap ($\sim$500 K) induced by the moir\'e superlattice so that we can neglect all the higher bands.  In Fig.~\ref{Fig:vF}(b) we show the calculated TBG Dirac velocity denoted as $v_F^*$ compared with the standard graphene Fermi velocity $v_F =10^8$ cm/s.   In Fig.~\ref{Fig:rho}, we show the calculated $ \rho (T, n, \theta)$ as a function of temperature for various angles and for various doping densities.  We also show the calculated $d\rho/dT$ as a function of the twist angle. $d\rho/dT$ reaches $\sim$100 ohms/K  to be compared with the value of 0.1 ohm/K for ordinary monolayer graphene at high temperatures.\cite{Hwang2008, Kim2010, Min2011} In fact, the calculated $d\rho/dT$ from Eq.~(\ref{rho_TBG}) keeps on increasing monotonically with decreasing twist angle before reaching $\theta_{\text{magic}}$, but Eq.~(\ref{rho_TBG})  is no longer applicable for $v_F^*<v_{ph}$.  The density dependence of our results is generally weak, and in fact, at high enough temperatures, $T>T_{BG}$, the density dependence essentially disappears.  The weak (very strong) dependence of the linear-in-$T$ resistivity on density (twist angle) is a specific verifiable prediction of our theory.  The density dependence is not very weak at low temperatures since $T_{BG} \sim \sqrt{n}$, and the true density-independent linear-in-$T$ behavior of $\rho(T)$ is only apparent for $T>T_L \sim T_{BG}/4$.  We emphasize that our results are obtained without any free adjustable parameters using the known values for graphene electron-phonon coupling: $D$=25 eV, $\rho_m=7.6\times 10^{-8}$ g/cm${^2}$, $v_{ph}=2\times 10^6$ cm/s.  There is some disagreement in the literature about the precise value of the graphene deformation potential constant with $D$=10$\sim$40 eV being quoted \cite{Guinea2007, Vasko2007, Rice1980, Woods2000, Suzuura2002, Pennington2003, Sugihara1983, Ono1966}.  Our results can be modified to take into account a different deformation potential constant simply by multiplying the calculated resistivity by $(D/25)^2$. 
We emphasize that we do not take into account the possibility of any modification in the phonon properties (e.g. $v_{ph}$ or $D$) due to the moir\'e superlattice structure-- any such modification would change the quantitative values of the resistivity, but not the trends as a function of $T$, $n$, and $\theta$.

We can see from Fig.~\ref{Fig:rho} that $\rho (T, n, \theta)$ could be very large, manifesting a linear dependence on $T$, a weak dependence on $n$, and a strong nonlinear dependence on $\theta$ around the magic angle [$\sim 1/(\theta-\theta_{\text{magic}})^2$ for $\theta>1.1^{\circ}$]. In addition, there is an implied dependence on the deformation potential constant $D$; if the actual value of D is 30 eV (rather than 25 eV as used in our theory), then the resistivity increases by another factor of 1.44.  Motivated by the holographic concept of Planckian dissipation which has recently been much discussed in the context of strange metals \cite{legros2018universal, zaanen2018planckian, zhang2018thermal, lucas2018operator,Hwang2018}, with the scattering rate becoming scale independent and being defined simply by $k_BT$, we show in Fig.~\ref{Fig:tau} our calculated scattering rate as a function of temperature for fixed values of density and twist angle.  We note that for the angle values used for these figures ($\theta =1.1^\circ$ to  $1.3^\circ$) and for the density ($n= 2 \sim 4 \times 10^{11}$ cm$^{-2}$) and temperature ($T$ up to 30K) ranges considered, the scattering rate obeys the Planckian bound being well below $T$. This is, however, a simple manifestation of the precise values of $D$ and $v_{ph}$ used in the calculation-- if $D$ ($v_{ph}$) is increased (decreased) by a factor of $2-3$, the Planckian bound is violated.

We now discuss our theory in the context of a comparison with the recently available TBG transport data \cite{MIT2018_rho,Columbia2018_rho} so that future theories could improve upon our approximations in achieving better agreement with experiments.  The theory makes four specific qualitative predictions for $\rho(n,T,\theta)$: (1) For $T>T_L$, where $T_L$ is a density dependent characteristic low temperature, the TBG resistivity shows very strong $\theta$-dependence; (2) the resistivity is linear in $T$ and is  very large for small $\theta$, almost diverging as $\theta$ approaches the magic angle $\theta_{\text{magic}}$; $d\rho/dT$ for $\theta$ close to $\theta_{\text{magic}}$ approaches or exceeds 100 ohm/K compared with the value of $\sim$ 0.1 ohm/K for the corresponding monolayer graphene; (3) the high-$T$ value of $d\rho/dT$ in the linear regime is essentially independent of the carrier density $n$; and, (4) $T_L$, down to which the linearity persists, goes as ~$\sqrt{n}$, and is approximately given by $T_L \sim 3.5 \sqrt{N} $ K where $N$ is the density $n$ measured in unit of $10^{11} \text{cm }^{-2}$. We note that the resistivity remains metallic, i.e. $d\rho/dT >0$, for $T<T_L$ down to a lower temperature $T_m$ ($<T_L$) both in our theory and in experiment, but the resistivity ceases to be linear in $T$ for $T<T_L$.

All four of these falsifiable predictions are in excellent qualitative agreement with the available TBG transport experimental data \cite{MIT2018_rho,Columbia2018_rho}.  Given that ours is the first and (so far) the only theory for the TBG metallic transport properties, such a qualitative agreement is all we can expect in this early stage of the subject where very little is understood or agreed upon about the underlying physics of the recent experimental observations.\cite{Cao2018Magnetic,Cao2018Super,Dean2018Tuning}  In particular, the strong (weak) angle (density) dependence of the linear-in- $T$ resistivity and the huge enhancement of $d\rho/dT$ close to magic angle follow naturally from our theory, and are rather non-trivial theoretical predictions which do not follow naturally from other mechanisms such as electron-electron interaction which should show a density dependence as it  scales on dimensional ground as $T/T_F$, where $T_F$, the Fermi temperature, manifests a $\sqrt{n}$ density dependence in graphene.

\begin{figure}[t]
	\includegraphics[width=1.0\columnwidth]{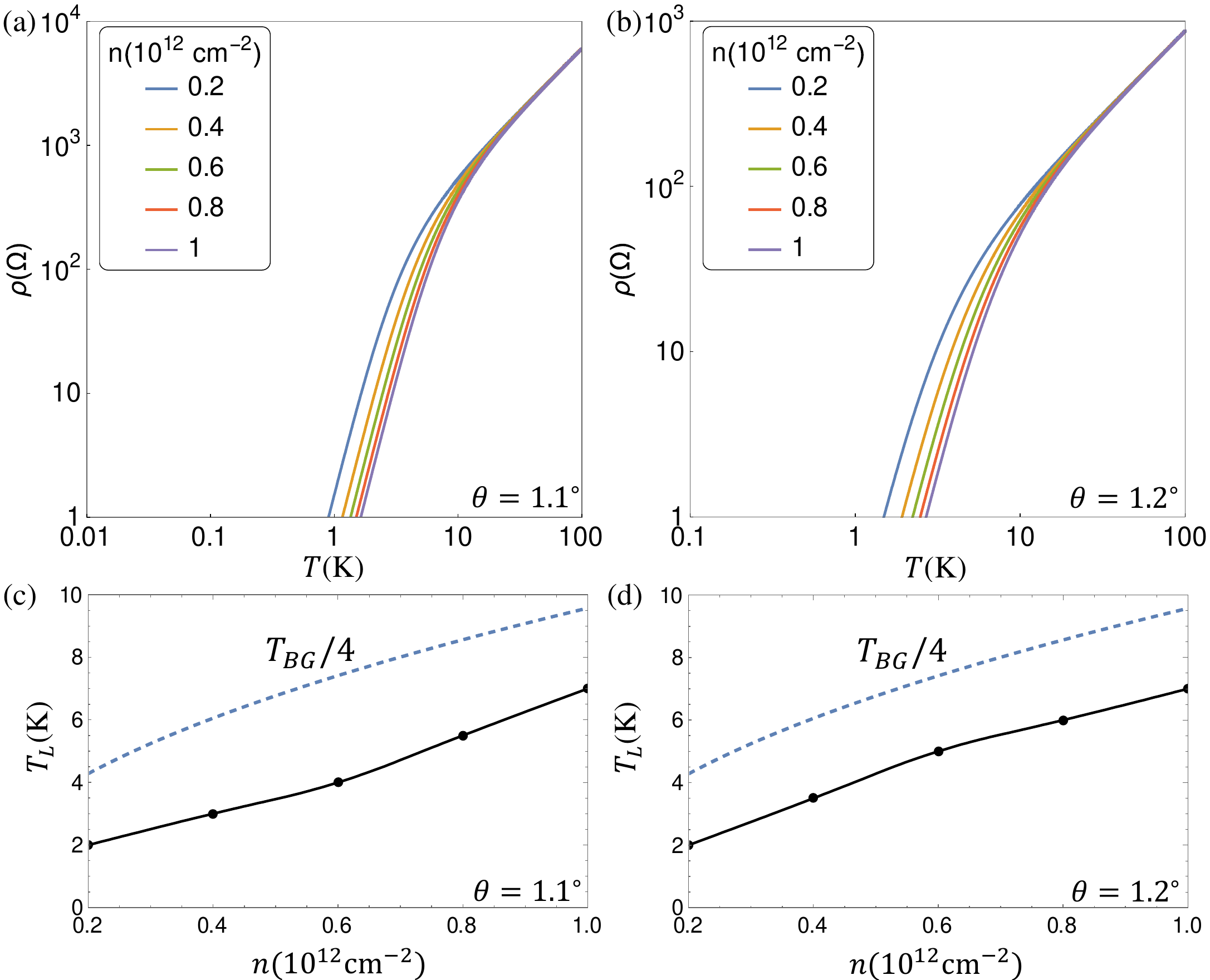}
	\caption{(a) and (b) Log-log plot of $\rho(T)$ in TBG. At high temperature $T>T_L$, $\rho$ is linear in $T$ and essentially independent of density $n$. At low temperature $T<T_L$,  $\rho$ scales as  $T^4$. (c) and (d) The crossover temperature $T_L$, down to which linearity in $\rho(T)$ persists, as a function of density $n$. The dashed lines show $T_{BG}/4$, while points on the solid lines represent $T_L$ read off from Figs.~\ref{rho_TBG}(c, d). $T_{BG}/4$ provides an upper bound on $T_L$.}
	\label{Fig:TBG}
\end{figure}

In Fig.~\ref{Fig:TBG}, we clearly bring out the density independence of the high-$T$ transport in our theory by showing results on a logarithmic scale for two twist angles, where $\rho(T)$ at high $T$ is manifestly density independent.  The linearity persists down to a temperature $T_L \sim T_{BG}/4$, which ranges around $2-8$ K for $n \sim 10^{11}-10^{12}$ cm$^{-2}$.  Both the density independence of the linear-in-$T$ resistivity regime and the persistence of linearity down to low temperatures are in good qualitative agreement with experiment.  We note that the log-log plots shown in Fig.~\ref{Fig:TBG} indicate that the low-$T$ behavior of $\rho(T,n)$ for $T<T_L$ is a $T^4$ Bloch-Gr\"uneisen resistivity due to suppressed phonon scattering.  It would be interesting to look for this low-temperature $T^4$ dependence of $\rho(T)$, but the samples must be extremely clean to see the low-temperature Bloch-Gr\"uneisen behavior as was done in Ref.~\cite{Kim2010} for monolayer graphene following the prediction of Ref.~\cite{Hwang2008}.

\begin{figure}[t]
	\includegraphics[width=1.0\columnwidth]{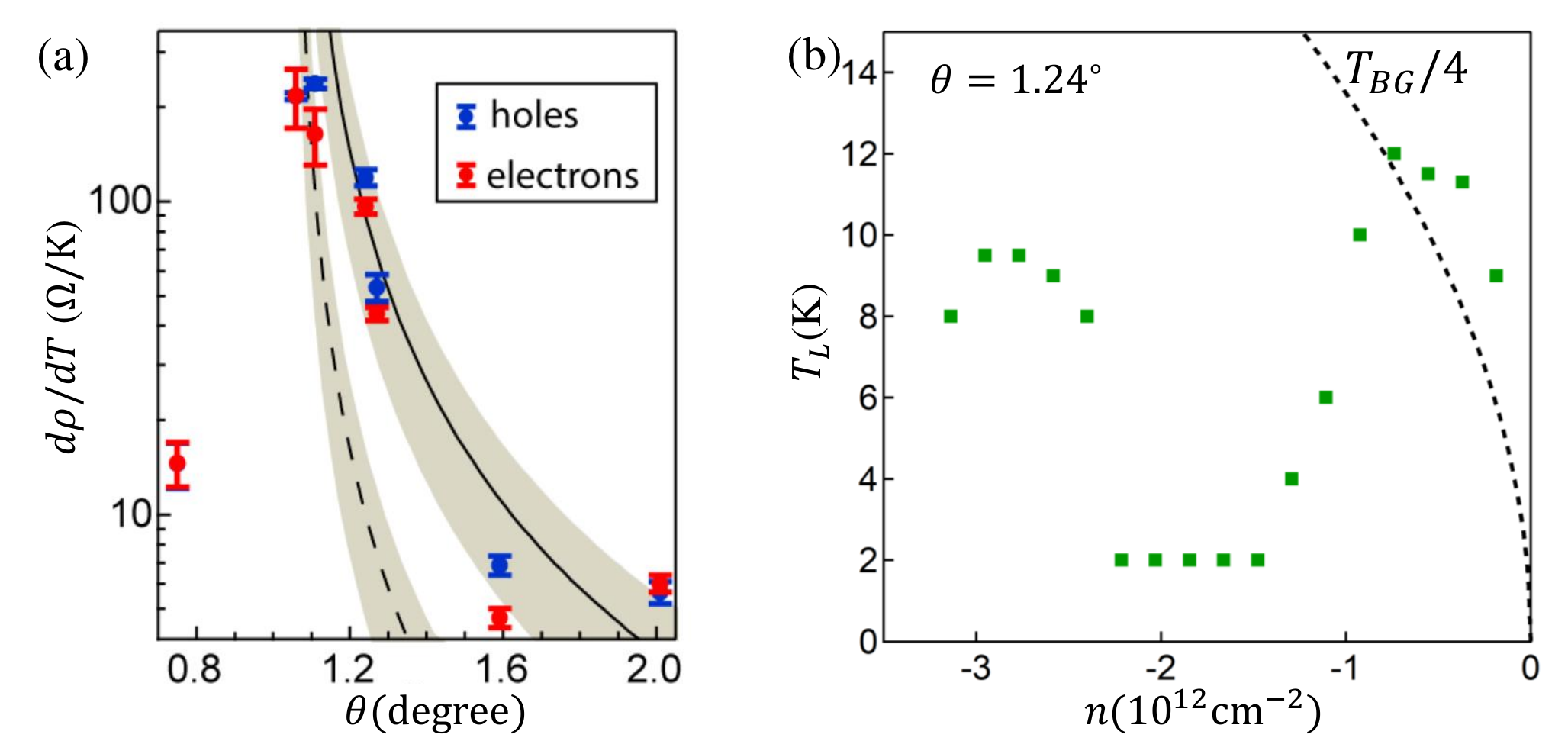}
	\caption{(a)$d\rho/d T$ in the linear regime as a function of twist angle. The markers are experimental data of Ref.~\cite{Columbia2018_rho} for different TBG devices. The dashed line shows the theoretical value $d\rho/d T$ based on Eq.~(\ref{rho_TBG}) with $D/v_{ph}$ set to its value in monolayer graphene. The solid line shows the theoretical prediction with  $D/v_{ph}$ enhanced by a factor of 3. This figure is adapted from Ref.~\cite{Columbia2018_rho}. (b) The crossover temperature $T_L$, down to which  $\rho(T)$ is linear in $T$, as a function of density $n$.  The dots are experimental data extracted from Ref.~\cite{Columbia2018_rho}. The dashed line represents the theoretical estimation with $T_L = T_{BG}/4$. The van Hove singularity occurring $\sim 2\times 10^{12} \text{cm}^{-2}$ in TBG is not included in our continuum theory as discussed in the text.}
	\label{Fig:exp}
\end{figure}

We emphasize that although we obtain excellent qualitative and perhaps semiquantitative agreement with the experimental results, our precise quantitative agreement is  modest.  In particular, the extracted $d\rho/dT$ from the experiment shows a quantitative disagreement with our theory using MLG phonon parameters.  This is shown in Fig.~\ref{Fig:exp}(a).
Typically, the observed $d\rho/dT$ is larger than our calculated values.  We can make experiment\cite{Columbia2018_rho} and our theory agree quantitatively if we use $D/v_{ph}$ appearing in the theory as a free parameter and adjust it by a factor of $2\sim3$.  While this agreement by adjusting $D/v_{ph}$ with a factor of $2\sim3$ is shown explicitly for a comparison with the experimental data of Ref.~\cite{Columbia2018_rho} in Fig.~\ref{Fig:exp}(a), we have verified the same to be true for the data of Ref.~\cite{MIT2018_rho} also.

A more important disagreement between our theory and experiment [as shown in Fig.~\ref{Fig:exp}(b)] is the behavior of $T_L$, the crossover temperature down to which the linear-in-$T$ resistivity behavior persists.  Although there is reasonable agreement between theory and experiment in the absolute magnitude of $T_L$ for $ n<10^{12}$ cm$^{-2}$, this agrement breaks down for $ n\sim 2\times10^{12}$ cm$^{-2}$ where the experimental $T_L$ goes down instead of increasing as $T_{BG}~\sqrt{n}$ as theory predicts.  The reason for this disagreement is clear:  There is a Fermi surface reconstruction for  $ n\sim 2\times10^{12}$ cm$^{-2}$ in the system (most likely Lifshitz transition of the Fermi surface associated with the van Hove singularity in the band structure) which is not included in our continuum theory.  This Fermi surface reconstruction makes the theory inapplicable at low temperatures near the van Hove singularity in the spectrum, making the experimental $T_L$ to lie below the theoretical $T_L$.  The interesting point, however, is that the theory still continues to give a good description of $\rho (T,n)$ at higher temperatures except that the linearity in $\rho(T)$ now persists to even lower temperatures than the theoretically predicted $T_L$.  This is understandable based on the fact that $T_L\sim T_{BG}\sim k_F$, and the Fermi surface reconstruction reduces the effective value of $k_F$, thus suppressing $T_L$ making the $T$-linear regime persist down to lower temperatures.  The persistence of the linear regime to lower temperatures than our theoretically predicted $T_L$ around the van Hove singularity is currently the most significant discrepancy between theory and experiment.\cite{MIT2018_rho, Columbia2018_rho}  We believe that this arises from the reduced effective value of $k_F$ in the vicinity of the van Hove singularity due to the Fermi surface reconstruction. 
Future theories going beyond our continuum theory should include the van Hove singularity explicitly, but such a theory will be numerical, losing the transparency of our theory.

What could cause the factor of 2 discrepancy between the theoretical and experimental values of $D/v_{ph}$ in demanding the precise quantitative agreement between the results of Refs.\cite{MIT2018_rho,Columbia2018_rho} and our phonon theory?  One possibility is that the realistic TBG phonon parameters are affected by the moir\'e superlattice effect, effectively raising $D/v_{ph}$ by a factor of 2 with respect to the phonon parameters of the simple monolayer graphene.  The other possibility is that other electron-phonon scattering processes and other phonon modes also contribute to $\rho(T)$, thus effectively increasing the relevant value of the parameter $D/v_{ph}$.  At this early stage of the subject, we judge that a quantitatively precise agreement between theory and a rapidly varying (as a function of the twist angle) experimental resistivity is beyond reach, and we note that our calculated resistivity should be taken as a lower bound for the experimental resistivity.  We can also think of the theory as having one free phonon parameter $D/v_{ph}$, which should be extracted experimentally.  The fact that this extracted parameter is within a factor of 2 of its value in monolayer graphene gives us confidence that the theory is a reasonable zeroth order approximation for the TBG high-temperature  transport properties.  We point out in this context that the twist angle for the actual experimental sample is neither precisely known nor a fixed quantity as it varies somewhat over the sample.\cite{kerelsky2018magic}  Since $d\rho/dT$ is a rapidly varying function of the twist angle $\theta$, a slight shift in the applicable $\theta$ values could provide a quantitative agreement between theory and experiment even without any adjustment of $D/v_{ph}$ for some twist angles. 

We note that, although we do not show it explicitly,  the inclusion of phonon umklapp scattering  does not change our results since umklapp scattering invariably involves phonons with momentum outside of the first moir\'e Brillouin zone and hence with relatively high energy, which is not particularly relevant for our results at low temperature.  Thus, our theory is correct to the extent the approximate flatband Dirac model is applicable.  In fact, even a departure from the Dirac model does not affect the results much since all it does is to impart a weak density dependence to $v_F^*$, but the dominant physics of its strong suppression remains valid.  

Finally, we comment on the experimentally observed decrease of $\rho (T)$ for much higher $T$ ($>50-100$K).  This has been explained\cite{Columbia2018_rho} as arising from the thermal activation of the electrons into higher bands which are not included in our theory.  It turns out that these higher bands are within few tens meV of the Fermi level, and hence thermal excitation into higher bands, which have much higher mobility leads to decreasing resistivity at higher temperatures. Thus, phonon induced linear-in-$T$ resistivity with a large density independent $d\rho/dT$ is strictly an intermediate temperature  property of TBG.

\section{Acoustic-phonon-mediated superconductivity}
\label{sec:SC}
We study superconductivity mediated by acoustic phonons in TBG, and show that these phonon modes can generate effective attractive interactions in four distinct pairing channels with respectively $s$-, $p$-, $d$- and $f$-wave symmetries. 
We emphasize that we include in the superconducting theory exactly the same in-plane acoustic phonon modes which we consider in Sec.~\ref{sec:resistivity} as causing the resistive scattering leading to the linear-in-$T$ resistivity.

In TBG, the in-plane acoustic modes associated with each individual graphene layer are combined into the layer symmetric and antisymmetric modes. A rigid in-plane translation of both layers leads to a global translation of the full system, while a rigid in-plane translation of one layer relative to the other layer generates a global shift of the moir\'e pattern in TBG. Therefore, both the layer symmetric and antisymmetric phonon modes remain gapless. This is to be contrasted with AB-stacked (Bernal) bilayer graphene, where the layer antisymmetric modes become gapped.

We make the assumption that the layer symmetric and antisymmetric modes with longitudinal (transverse) displacement have the same phonon velocity, because interlayer phonon coupling is weak. Under this assumption, it is more convenient to use the acoustic modes of individual graphene layer,  and the electron-phonon coupling Hamiltonian is then given by:
\begin{equation}
H_{\text{EPC}}=\frac{D}{\sqrt{\mathcal{A}}}\sum_{\qq, \ell} \sqrt{\frac{\hbar }{2\rho_m \omega_{\qq}}}(-i \qq\cdot\hat{e}_{\qq})(a_{\qq,\ell}+a_{-\qq,\ell}^\dagger) \hat{n}_{\qq, \ell},
\label{HEPC}
\end{equation}
where $a_{\qq,\ell}$ represents the longitudinal acoustic  phonon operator with layer index $\ell$, frequency  $\omega_{\qq} = v_{ph} |\qq|$, and displacement unit vector  $\hat{e}_{\qq}$. In Eq.~(\ref{HEPC}), the electron density operator $\hat{n}_{\qq,\ell}$ is defined as  $\sum_{\kk \xi \sigma s} c_{(\kk+\qq) \xi\sigma \ell s}^\dagger c_{\kk \xi \sigma \ell s}$, where $c_{\kk \xi \sigma \ell s}$ is the electron operator with valley ($\xi=\pm$), sublattice ($\sigma=A, B$), layer ($\ell=t, b$) and spin $(s=\uparrow,\downarrow)$ indices, and momentum label $\kk$ measured relative to the Dirac point in valley  $\xi$ and layer $\ell$. The meaning of other terms in Eq.~(\ref{HEPC}) are as follows: $\mathcal{A}$ is the system area, $D$ is the deformation potential, and $\rho_m$ is the mass density. Note that Eq.~(\ref{HEPC}) is exactly the same Hamiltonian used in Sec.~\ref{sec:resistivity} for calculating the electron-phonon interaction induced resistive scattering.

In the Hamiltonian (\ref{HEPC}), we take into account the modulation of on-site potential due to acoustic phonons, but neglect their effect on the interlayer tunneling \cite{Lian2018twisted}; we speculate that the latter effect is weaker due to the smallness of interlayer tunneling parameter ($\sim$ 0.1 eV) compared to the deformation potential $D$ ($\sim$ 25 eV), although a careful study remains to be performed for a direct comparison between different electron-phonon scattering processes. The current work focuses on the well-defined model of  Eq.~(\ref{HEPC}) to search for the superconducting instability.

By integrating out the phonon modes, we obtain the following phonon-mediated dynamical electron-electron interaction Hamiltonian:
\begin{equation}
H_{\text{att}}=-\frac{1}{\mathcal{A}} \frac{D^2}{2\rho_m v_{ph}^2}\sum_{\qq,\omega_n \ell}
\frac{\omega_{\qq}^2}{\omega_{\qq}^2+\omega_n^2}
\hat{n}_{\qq,\ell} \hat{n}_{-\qq,\ell},
\end{equation}
where $\omega_n$ is the bosonic Matsubara frequency. To make an order-of-magnitude estimation of the superconducting critical temperature, we neglect the frequency dependence in the attractive interaction and obtain the following Bardeen-Cooper-Schrieffer (BCS) pairing Hamiltonian:  
\begin{equation}
H_{\text{BCS}}= -4 g_0 \sum_{\ell,\sigma,\sigma',s,s'}\int d\rr \hat{\psi}_{+ \sigma \ell s}^\dagger  \hat{\psi}_{- \sigma' \ell s'}^\dagger
\hat{\psi}_{- \sigma' \ell s'} \hat{\psi}_{+ \sigma \ell s},
\label{HBCS}
\end{equation}
where the attractive interaction strength is $g_0 = D^2/(4 \rho_m v_{ph}^2) \approx 82.3 $ meV nm$^2$, $\hat{\psi}_{\xi \sigma \ell s}$ is the  real-space electron operator, and all $\hat{\psi}$ and $\hat{\psi}^{\dagger}$ operators are understood to be at the same coarse-grained position $\rr$. 
In Eq.~(\ref{HBCS}), we have only retained terms that pair electrons from opposite valleys \cite{Wu2018phonon}.

As a side remark, we note that the experimentally measured  coherence length for the superconducting state in a twisted bilayer graphene device with 1.16$^{\circ}$ twist angle is about 52 nm \cite{Cao2018Super}, while the corresponding moir\'e period is about 12 nm. Therefore, the measured coherence length is four times as large as the moir\'e period, which indicates that the superconductivity in twisted bilayer graphene is not disordered on the moir\'e scale and also provides a justification for our BCS mean-field theoretical treatment. The superconducting coherence length can be theoretically estimated as $\hbar v_F^* /(\pi \Delta_{\text{SC}})$, where $ v_F^*$ is the renormalized velocity of the nearly flat bands and  $\Delta_{\text{SC}}$ is
the superconducting gap. If we take $ v_F^*$ to be $4 \times 10^{4} $ m/s (4\% of the monolayer Dirac velocity), and $\Delta_{\text{SC}}$ to be 1 K  ($\sim$0.086 meV), then the theoretical coherence length is about 97 nm, which is on the same order as the experimentally measured coherence length. This consistency indicates that the Cooper
pairing indeed forms between electrons in the flat bands. These estimates for the coherence length as compared with the unit cell size of the moire superlattice also show that the mean field theory is a reasonably good approximation for treating superconductivity in TBG systems as carried out in our theory.

In $H_{\text{BCS}}$, there are four different pairing channels that are distinguished by their orbital and spin characters: 
(1) $s$-wave with intrasublattice spin-singlet pairing, i.e., $\epsilon_{s s'} \hat{\psi}_{+ \sigma \ell s}^\dagger  \hat{\psi}_{-  \sigma \ell s'}^\dagger $, where $\epsilon = i s_y $ is the fully antisymmetric tensor; 
(2) $p$-wave with intersublattice spin-triplet pairing, e.g., $\mathcal{F}_{s s'} \hat{\psi}_{+ A \ell s}^\dagger  \hat{\psi}_{- B \ell s'}^\dagger $, where $\mathcal{F}$ can be any one of the three symmetric tensors $(s_0 \pm s_z)/2$ and  $s_x$; 
(3) $d$-wave with intersublattice spin-singlet pairing, e.g., $\epsilon_{s s'} \hat{\psi}_{+ A \ell s}^\dagger  \hat{\psi}_{- B \ell s'}^\dagger $; 
and (4) $f$-wave with intrasublattice spin-triplet pairing, i.e., $\mathcal{F}_{s s'} \hat{\psi}_{+ \sigma \ell s}^\dagger  \hat{\psi}_{- \sigma \ell s'}^\dagger $.
Here $s$ and $f$ pairings are only distinguished by the spin part, and similarly for $p$ and $d$ pairings. Because of the sublattice pseudospin chirality
under the threefold rotation,
$\hat{C}_3 \hat{\psi}^{\dagger}(\rr) \hat{C}_3^{-1} = \exp[i 2\pi \sigma_z \xi_z /3] \hat{\psi}^{\dagger}(\mathcal{R}_3 \rr) $, intersublattice Cooper pairs carry a finite orbital angular momentum \cite{Wu2018phonon} and therefore belong to $p$- or $d$-wave pairings.

\begin{figure}[t]
	\includegraphics[width=1.0\columnwidth]{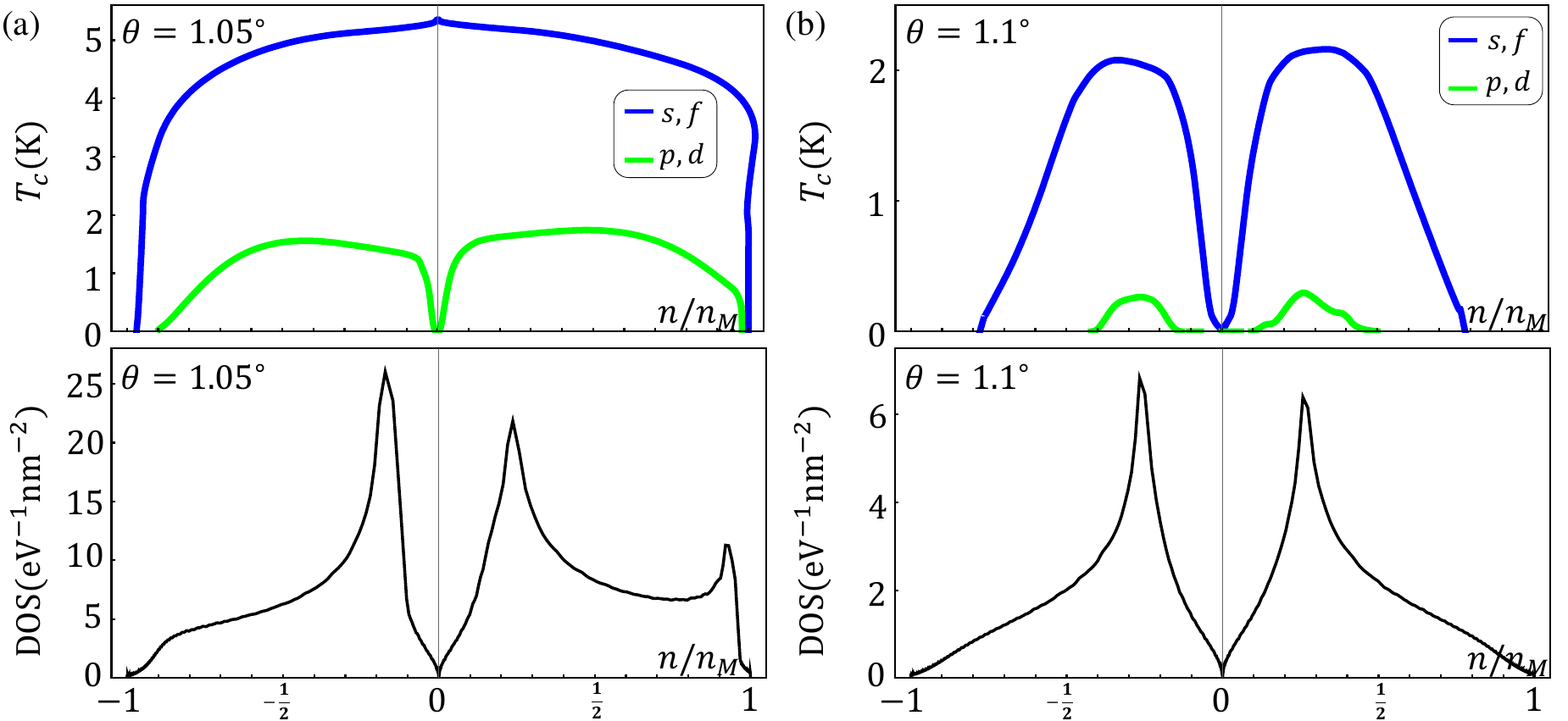}
	\caption{Superconducting critical temperature $T_c$ in different channels (upper panel), and the DOS per spin and per valley as a function of filling factor $n/n_M$ (lower panel). $n/n_M$ is 1 when the flatbands are fully filled. The twist angle is $1.05^{\circ}$ in (a) and $1.1^{\circ}$ in (b). }
	\label{Fig:Tc}
\end{figure}

The pairing Hamiltonian $H_{\text{BCS}}$ in Eq.~(\ref{HBCS}) combined with the single-particle moir\'e Hamiltonian  has  spin SU(2)$\times$SU(2) symmetry, that is, each valley has its own spin rotational symmetry. This enlarged symmetry enforces the degeneracy between intervalley spin singlet and spin triplet pairings, i.e., $s$- and $f$-wave have degenerate critical temperature $T_c$, and the same is true for $p$- and $d$-wave within our model. This symmetry has also been noticed in Refs.~\cite{You2018,Isobe2018}, which study  Coulomb repulsion mechanisms for superconductivity.
We note that any mechanism, either extrinsic or spontaneous, breaking this enlarged spin symmetry, will lift the degeneracy between singlet and triplet channels.

We calculate $T_c$ separately for each channel by solving the corresponding linearized gap equation with procedures outlined in Refs.~\cite{Wu2018phonon,wu2018topological}. Details of the theoretical analysis are given in Appendix~\ref{AppC}. We note that the pair amplitude is identical in the two layers in the $s$- and $f$-wave channels but phase shifted by $\pi$ in the $p$- and $d$-wave channels  \cite{Wu2018phonon}. Fig.~\ref{Fig:Tc} shows the calculated $T_c$ as a function of the moir\'e band filling factor for two representative twist angles close to the magic angle. Because of the strongly enhanced density of states (DOS) for the flatbands near magic angle, $T_c$ in all channels can reach order of 1$\sim$5 K, with $s$- and $f$-wave channels being the leading instability. 

The dimensionless BCS electron-phonon coupling constant is $\lambda^* = g_0 N(\varepsilon_F)$, where $N(\varepsilon_F)$ is the DOS per spin and per valley at the Fermi energy $\varepsilon_F$. Typical values of $\lambda^*$ for the DOS shown in Fig.~\ref{Fig:Tc} are on the order of 0.5, indicating strong electron-phonon coupling. This should be contrasted with the corresponding dimensionless coupling constant for electron-acoustic phonon interaction in monolayer graphene which is $\sim 0.0001 $ for electron density $n\sim 10^{12} \text{cm}^{-2}$.

We note that another dimensionless electron-phonon coupling constant $\lambda_{tr}$  can be extracted from the transport scattering rate  \cite{Hwang2018}:
\begin{equation}
\hbar/\langle \tau \rangle=2\pi \lambda_{tr} k_B T.
\end{equation}
where $\hbar/\langle \tau \rangle$, as shown in in Fig.~\ref{Fig:tau}, is obtained from the resistivity calculation. Remarkably,   $\lambda^*$ and $\lambda_{tr}$ are proportional to each other with the simple relation $\lambda^* = 2 \lambda_{tr} /F(\theta) $, where $F(\theta)$ is the form factor (arising from moir\'e wave functions) that enters into the resistivity calculation [Eq.~(\ref{rho_TBG})] but not into the superconducting susceptibility calculation. Therefore, the dimensionless electron-phonon coupling for the low-temperature superconductivity  and for the  high temperature transport scattering rate  should be connected to each other since both physics is controlled by the same electron-phonon coupling. 
The interesting new physics here is that, unlike in ordinary metals where $\lambda_{tr}$ and $\lambda^*$ are typically within 10\% of each other\cite{PAllen}, in TBG the two couplings, although related, differ by a numerical factor depending on the twist angle.  Close to the magic angle they differ roughly by a factor of 4. In monolayer graphene, $\lambda^* = 2 \lambda_{tr}$, where the prefactor $2$ arises because the Dirac fermion matrix elements have different forms in the scattering problem and the BCS pairing problem.  
The flatband graphene velocity suppression leads to a giant linear-in-$T$ resistivity at higher temperatures with the slope $d\rho/dT$ determined by  the same enhanced electron-phonon coupling that leads to the superconducting instability at low temperatures.
We emphasize, however, that our phonon scattering theory for the TBG strange metal transport should apply independent of the mechanism for superconductivity since the former applies in the metallic regime well above $T_c$, where phonon scattering is always present. Thus, even if some other mechanism supersedes to induce superconductivity in the system independent of electron-phonon interaction in the ground state, high-temperature resistivity should still be dominated (to the extent at least of a lower bound) by our theory of Sec.~\ref{sec:resistivity}.

Finally, we discuss the role played by Coulomb repulsion in the low-temperature phase diagram. As is well known, the Coulomb repulsion can suppress the phonon-induced superconducting instability (the so-called $\mu^*$ effect where Coulomb repulsion reduces the effective electron-electron attractive interaction induced by electron-phonon interactions). However, the attraction generated by electron-acoustic phonon coupling could overcome Coulomb repulsion through retardation effects. A quantitative theory of the transition temperature $T_c$ in TBG that includes Coulomb repulsion and retardation effects remains to be developed. The net effect of Coulomb repulsion would be to quantitatively suppress the superconducting $T_c$ calculated in Fig.~\ref{Fig:Tc}, and this suppression is likely to be dependent on both twist angle ($\theta$) and density ($n$). The Coulomb repulsion can also drive correlated insulating states at commensurate fillings with integer number of electrons or holes per moir\'e unit cell. Here our assumption is that the electron-phonon coupling and electron-electron Coulomb repulsion compete with each other, and respectively give rise to the superconducting and correlated insulating states. At commensurate fillings (1/4, 1/2, 3/4, etc.) an insulator develops suppressing superconductivity completely whereas at fillings away from these commensurate fillings, the system can manifest superconductivity induced by the strong electron-phonon interaction. We speculate that in the absence of the insulating states, phonon-induced superconductivity could exist throughout the moir\'e flatbands for twist angles close to the magic angle because of the narrow bandwidth and the giant enhancement of the electron-acoustic phonon coupling.  Note that other phonon modes, considered in other works\cite{Wu2018phonon,wu2018topological}, could also produce a finite $T_c$ in the system.

The spin SU(2)$\times$SU(2) symmetry in the pairing Hamiltonian (\ref{HBCS}) can be broken down to spin SU(2) symmetry by atomic scale interactions, and therefore, the degeneracy between spin singlet and triplet pairings can be lifted. For example, atomic scale on-site Coulomb repulsion will  suppress the $s$-wave channel but not  other channels. As shown in Ref.~\cite{Wu2018phonon}, intervalley optical phonons generate attractive interaction in $s$-wave and $d$-wave channels, but repulsive interaction in $p$-wave and $f$-wave channels. Therefore, the competition between different pairing channels depend on Coulomb repulsion and also other phonon modes.

\section{Conclusion}
\label{sec:Discussion}

In conclusion, we have shown that in principle it is possible for strange metallic behavior with a large linear-in-temperature resistivity to arise from perfectly ordinary electron-phonon interaction mechanisms within the Fermi liquid theory through flatband physics leading to a giant electron-phonon coupling. We emphasize that our transport theory uses the (moir\'e-modified) continuum Dirac Hamiltonian for low carrier density  and neglects interaction and disorder effects, and as such is an approximate theory, which should be thought of as a zeroth order theory for TBG transport involving only acoustic phonon scattering.  The fact that such a simple theory already accounts for all of the qualitative features of the experimental data is encouraging and noteworthy. It should be possible for future works to systematically improve the transport theory by including the effects left out in our  theory (e.g. more realistic band structure, inclusion of van Hove singularities, inclusion of phonon-induced inter-band scattering, higher band effects on transport, disorder effects), but we believe that our theory establishes the leading-order transport theory for twisted bilayer graphene as evidenced by the agreement of our theory with recent experiments. 

We have also shown that the same electron-phonon interaction, leading to a large linear-in-T resistivity at higher temperatures ($>5-10$ K), can also produce superconductivity with $T_c \sim 1 $ K in $s$, $p$, $d$ and $f$ orbital pairing channels.  The highly unusual exotic higher-order orbital pairing induced by electron-phonon interaction becomes possible in this system because of the sublattice pseudospin chirality and the enlarged spin rotational symmetry in TBG.  The superconducting $T_c$ is relatively high simply because the electron-phonon interaction is very strong due to the band flattening. Thus, the large linear-in-$T$ resistivity and the high superconducting transition temperature are connected although the dimensionless electron-phonon coupling parameters entering the two properties could differ by a factor of 2-4 due to the pseudospin properties of twisted bilayer graphene. We emphasize that our theory of phonon-induced TBG superconductivity uses the full moir\'e band structure incorporating physics beyond the effective Dirac model such as  the van Hove singularities in the density of states.

\section{Acknowledgment}
The authors thank Cory Dean (Columbia), Andrea Young (UCSB), and Pablo Jarillo-Herrero (MIT) for many useful discussions on their respective unpublished experimental data on TBG.  The authors particularly thank A. Young, C. Dean, M. Yankowitz, and H. Polshyn  for  sharing their voluminous experimental data before publication. F. W. also thanks B. Lian and M. Zaletel for stimulating discussions. This work is supported by Laboratory for Physical Sciences.

\appendix

\section{Moir\'e band structure and form factor}
\label{AppA}
\begin{figure}[t]
	\includegraphics[width=0.9\columnwidth]{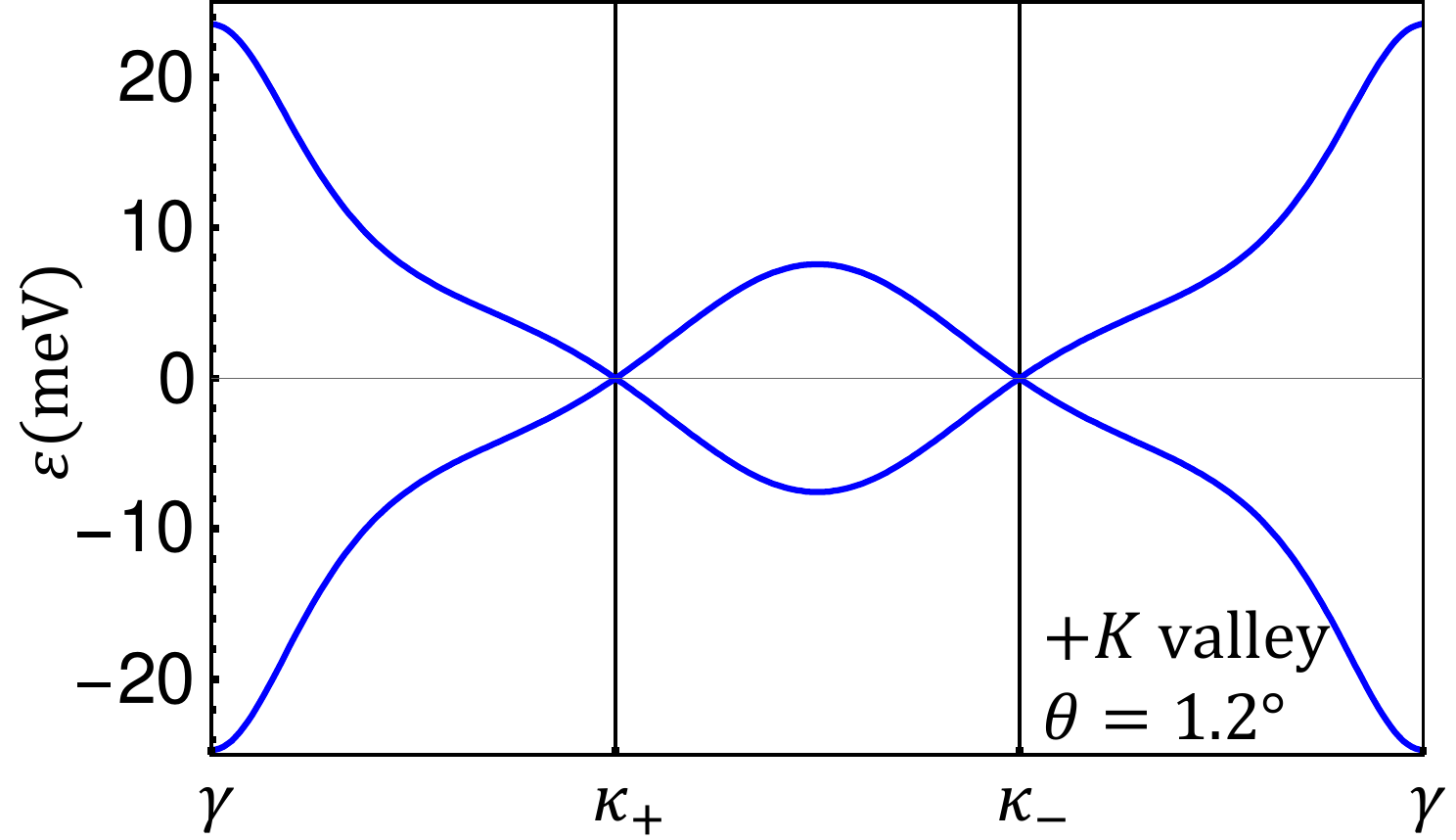}
	\caption{Moir\'e band structure along high symmetry directions in the moir\'e Brillouin zone for $+K$ valley and twist angle $\theta=1.2^{\circ}$. Dirac cones are centered at the moir\'e Brillouin zone corners $\kappa_{\pm}$ }
	\label{Fig:moire_band}
\end{figure}

We have used the moir\'e Hamiltonian of Ref.~\cite{Bistritzer2011} to calculate the moir\'e band structure. The moir\'e Hamiltonian in valley $+K$ is given by:
\begin{equation}
\mathcal{H}_{+}=\begin{pmatrix}
h_{\mathfrak{b}}(\kk) & T(\rr) \\
T^{\dagger}(\rr) & h_{\mathfrak{t}}(\kk)
\end{pmatrix}.
\label{Hmoire}
\end{equation}
Here $h_{\ell}(\kk)$ is the Dirac Hamiltonian of layer $\ell$,
\begin{equation}
h_{\ell}(\kk) = e^{-i\ell \frac{\theta}{4} \sigma_z }[\hbar v_F (\kk-\boldsymbol{\kappa}_{\ell})\cdot \boldsymbol{\sigma}]e^{+i\ell \frac{\theta}{4} \sigma_z},
\end{equation}
where $\ell$ takes value of $+1$ and $-1$ respectively for the bottom ($\mathfrak{b}$) and top ($\mathfrak{t}$)  layer,
$v_F$ is the monolayer Dirac velocity($\sim 10^6$ m/s), and
$\boldsymbol{\sigma}$ are sublattice pseudospin Pauli matrices. Because of the layer relative rotation, the valley $+K$ Dirac cone position in layer $\ell$ is shifted to $\boldsymbol{\kappa}_{\ell}$. We choose $\boldsymbol{\kappa}_{\ell}$ to be $[4\pi/(3 a_M)](-\sqrt{3}/2, -\ell/2)$ , where $a_M$ is the moir\'e lattice constant.  The interlayer tunneling $T(\rr)$ are sublattice dependent and vary spatially with moir\'e period:
\begin{equation}
T(\rr)=
T_0
+e^{-i \bb_+ \cdot \rr} T_{+1}
+e^{-i \bb_- \cdot \rr} T_{-1},
\end{equation}
where the moir\'e reciprocal lattice vectors $\bb_{\pm}$ are equal to $[4\pi/(\sqrt{3} a_M)](\pm1/2, \sqrt{3}/2)$. The tunneling terms $T_0$ and $T_{\pm 1}$  are given by:
\begin{equation}
T_j = w_0 \sigma_0 + w_1[\cos(2\pi j/3)\sigma_x+\sin(2\pi j/3)\sigma_y]
\end{equation}
where $w_0$ and $w_1$ are two  parameters that respectively characterize the tunneling
in AA and AB/BA regions in the moir\'e pattern. Values of $w_0$ and $w_1$ are different because of the layer corrugation effect. We take $w_0$ = 90 meV and $w_1$ = 117 meV from Ref.~\cite{Jung2014}.
A representative moir\'e band structure is shown in Fig.~\ref{Fig:moire_band} for $\theta=1.2^{\circ}$.

We calculate the velocity $v_F^*$ at Dirac points as a function of twist angle using the moir\'e Hamiltonian (\ref{Hmoire}) with parameters given above, and the result is shown in Fig.~\ref{Fig:vF}. The Dirac velocity reaches a minimum
value at  angle  $\sim$ 1.025$^\circ$, which is the largest magic angle in our model calculation. In real systems, there may be twist angle variations and atomic relaxations, making a quantitative comparison between theory and experiment challenging.

We have used the full moir\'e band structure obtained from 
Hamiltonian (\ref{Hmoire}) to calculate the superconducting transition temperature. In the resistivity calculation,  the Dirac velocity $v_F^*$, as shown in Fig.~\ref{Fig:vF}, is also from the full moir\'e band structure calculation; the form factor $F(\theta)$ that accounts for the modification of the moir\'e wave function due to the interlayer tunneling, however, is calculated using a perturbation theory, which is a simplification in order to achieve analytical results. The full moir\'e band structure from Eq.~(\ref{Hmoire}) combined with the electron-phonon coupling Hamiltonian (\ref{HEPC}) can be used for a numerical study of phonon induced resistive scattering in TBG, which we leave for future work. Such a numerical study will be important in order to address detailed band structure effects, such as the influence of van Hove singularity on transport, but it should agree with our analytical theory at low carrier density. As we discuss in the main text, we believe that the experimental persistence of linearity in the resistivity to temperatures much lower than the nominal $T_{BG}/4$ around $n \sim 2\times 10^{12} \text{cm}^{-2}$ density most likely arises from the Fermi surface Lifshitz transition induced by the van Hove singularity leading to an effective reduction in the Fermi momentum.

To calculate $F(\theta)$, we use the perturbation theory of Ref.~\cite{Bistritzer2011}. This perturbation theory is justified for twist angle $\theta$ down to the largest magic angle, because it correctly predicts the existence and the value of this magic angle. In the perturbation theory, the two degenerate states at the Dirac point $\boldsymbol{\kappa}_{+1}$ are given by:
\begin{equation}
\begin{aligned}
&|\Psi_A\rangle=\frac{1}{\sqrt{1+\beta}}\{|\Psi_A^{(0)}\rangle, |\Psi_A^{(1)}\rangle, |\Psi_A^{(2)}\rangle, |\Psi_A^{(3)}\rangle \},\\
&|\Psi_B\rangle=\frac{1}{\sqrt{1+\beta}}\{ |\Psi_B^{(0)}\rangle, |\Psi_B^{(1)}\rangle, |\Psi_B^{(2)}\rangle, |\Psi_B^{(3)}\rangle \}.
\end{aligned}
\label{Psi_pert}
\end{equation}
Here $|\Psi_\sigma^{(0)}\rangle$ is the  wave function associated with layer $\ell=+1$ and momentum $\boldsymbol{\kappa}_{+1}$, while $|\Psi_\sigma^{(1, 2,3)}\rangle$ are wave functions associated with the opposite layer $\ell=-1$ and momentum $\boldsymbol{\kappa}_{+1}$, $\boldsymbol{\kappa}_{+1}+\bb_{+}$, $\boldsymbol{\kappa}_{+1}+\bb_{-}$, respectively. $|\Psi_A^{(0)}\rangle$ is $(1,0)^{\text{T}}$ , where the two components in the spinor represent $A$ and $B$ sublattices; correspondingly,  $|\Psi_B^{(0)}\rangle$ is $(0,1)^{\text{T}}$. Other spinors are specified as follows:
\begin{equation}
\begin{aligned}
|\Psi_A^{(n)}\rangle & = (- i \alpha_1, e^{i \phi_n} \alpha_0)^{\text{T}},\\
|\Psi_B^{(n)}\rangle & = (e^{-i \phi_n} \alpha_0, i \alpha_1)^{\text{T}},
\end{aligned}
\end{equation}
where $n$ is 1, 2 and 3, $\phi_n = (11- 8n)\pi/6 $, and $\alpha_j= w_j /(\hbar v_F |\boldsymbol{\kappa}_{+1}|)$.
In Eq.~(\ref{Psi_pert}), $\beta$ is equal to $3(\alpha_0^2+\alpha_1^2)$. $|\Psi_A\rangle$ and $|\Psi_B\rangle$ are dominated by states in layer $\ell=+1$ when the interlayer tunneling is weak ($\alpha_j \ll 1$), consistent with the fact that the Dirac cone at $\boldsymbol{\kappa}_{+1}$ originates from layer $\ell=+1$.

The Dirac cone centered at $\boldsymbol{\kappa}_{+1}$ can be described in the basis of  $|\Psi_A\rangle$ and $|\Psi_B\rangle$ using the Dirac Hamiltonian with the renormalized velocity $v_F^*$:
\begin{equation}
\mathcal{H}_{\boldsymbol{\kappa}_{+1}} = \hbar v_F^* \kk\cdot \boldsymbol{\sigma},
\label{Hkappa}
\end{equation}
where $\kk$ is measured relative to $\boldsymbol{\kappa}_{+1}$. The wave function for $\mathcal{H}_{\boldsymbol{\kappa}_{+1}} $ can be written as:
\begin{equation}
\begin{aligned}
|\zeta, \kk \rangle &= \frac{1}{\sqrt{2}} ( |\Psi_A\rangle  + \zeta e^{i \phi_{\kk}} |\Psi_B\rangle  ),\\
& =(|\zeta, \kk \rangle_0, |\zeta, \kk \rangle_1, |\zeta, \kk \rangle_2, |\zeta, \kk \rangle_3),\\
|\zeta, \kk \rangle_n &=  \frac{1}{\sqrt{2 (1+\beta )}} ( |\Psi_A^{(n)}\rangle  + \zeta e^{i \phi_{\kk}} |\Psi_B^{(n)}\rangle  ),
\end{aligned}
\end{equation}
where $\zeta = \pm 1$ is the band index, and $\phi_{\kk}$ is the orientation angle of momentum  $\kk$. $|\zeta, \kk \rangle_n$ is the wave function on layer $\ell=+1$ for $n=0$, but on the opposite layer  $\ell=-1$ for $n=$1, 2, and 3.

We use the electron-phonon coupling  in Eq.~(\ref{HEPC}) to study resistive scattering induced by phonons. The acoustic phonons associated with different layers lead to independent scattering processes. Therefore, the matrix element amplitude that enters into the scattering rate equation is:
\begin{equation}
\begin{aligned}
&|{}_0\langle \zeta', \kk' |\zeta, \kk \rangle_0|^2+|\sum_{n=1}^{3} {}_n\langle \zeta', \kk' |\zeta, \kk \rangle_n |^2\\
 = & F(\theta)(1+ \zeta \zeta' \cos \phi_{\kk \kk'})/2
\end{aligned},
\label{MEA}
\end{equation} 
where $F(\theta)$, equal to $(1+\beta^2)/(1+\beta)^2$, is the form factor, and $\phi_{\kk \kk'}$ is the scattering angle between $\kk$ and $\kk'$.  $F$ is between 0.5 and 1, and becomes 1  when the interlayer tunneling is zero. Near the magic angle, $F$ is close to 0.5.

\section{Transport theory}
\label{AppB}
We describe in detail a transport theory of the electronic resistivity $\rho$ in TBG due to the longitudinal acoustic phonon scattering. Equation (\ref{rho_TBG}) in the main text has been derived using the quasi-elastic approximation \cite{Hwang2008, Min2011, Hwang2018}. In this approximation, the relevant phonon energy is assumed to be much smaller than the Fermi energy, which is valid when the Dirac Fermi velocity ($v_F^*$) is much greater than the phonon velocity ($v_{ph}$). Here we present a theory that does not rely on this approximation, and we show that Eq.~(\ref{rho_TBG}) is accurate even when $v_F^*$ and $v_{ph}$ have comparable magnitude.

Because the low-energy bands in TBG have narrow bandwidth near the magic angle, both the conduction and valence bands can make significant contribution to the conductivity when the temperature exceeds the Fermi temperature. The conductivity ($\sigma_{tr}=1/\rho$) can be decomposed as:
\begin{equation}
\sigma_{tr} = \sum_{\zeta=\pm 1} \sigma_{\zeta},
\end{equation}
where $\sigma_{\zeta}$  is the conductivity contributed by the conduction ($\zeta=+1$) or valence ($\zeta=-1$) band. 

Within Boltzmann transport theory, the conductivity  $\sigma_{\zeta}$ can be calculated using the relaxation time approximation:
\begin{equation}
\sigma_{\zeta} = g_s g_v g_l e^2 \int \frac{d^2 \kk}{(2\pi)^2} \tau_{\zeta,\kk} \frac{v_F^{*2}}{2}\Big( - \frac{\partial f(\varepsilon_{\zeta, \kk})}{ \partial \varepsilon_{\zeta, \kk}} \Big),
\end{equation}
where $\varepsilon_{\zeta, \kk}=\zeta \hbar v_F^* |\kk|$, $f(\varepsilon_{\zeta, \kk})$ and $\tau_{\zeta,\kk}$  are respectively the energy, Fermi occupation number and relaxation time of the state in band $\zeta$ at momentum $\kk$.  

\begin{figure}[t]
	\includegraphics[width=1.0\columnwidth]{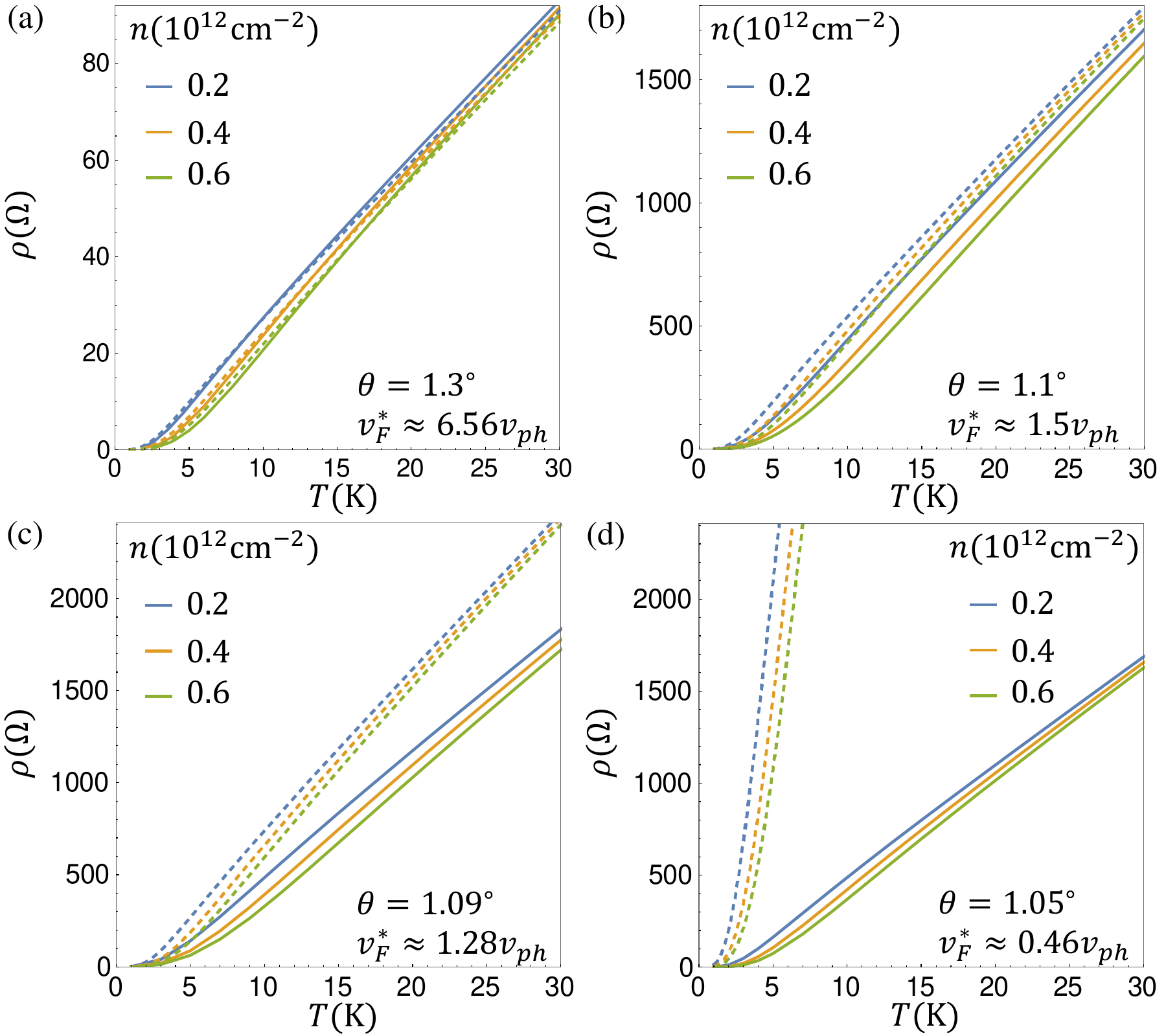}
	\caption{Resistivity $\rho$ as a function of temperature $T$. Twist angle $\theta$ is, respectively, $1.3^{\circ}$, $1.1^{\circ}$, $1.09^{\circ}$ and $1.05^{\circ}$ in (a), (b), (c) and (d). The Dirac velocity $v_F^*$ varies as a function of $\theta$. In each plot, solid lines are calculated based on Eq.~(\ref{rho_full}), while dashed lines are obtained from Eq.~(\ref{rho_TBG}). }
	\label{Fig:rho_compare}
\end{figure} 

The scattering rate can be expressed as:
\begin{equation}
\frac{1}{\tau_{\zeta,\kk}} =\sum_{\zeta',\kk'} (1-\zeta \zeta' \cos \phi_{\kk \kk'}) W_{\kk \kk'}^{\zeta \zeta'} \frac{1-f(\varepsilon_{\zeta', \kk'})}{1-f(\varepsilon_{\zeta, \kk})},
\end{equation}
where $\zeta'=\zeta$ ($\zeta'\neq \zeta$) terms represent intra (inter) band scattering processes.
By using the electron-phonon coupling in Eq.~(\ref{HEPC}) and the matrix element amplitude in Eq.~(\ref{MEA}), we obtain the transition rate $W_{\kk \kk'}^{\zeta \zeta'}$ of the form:
\begin{equation}
W_{\kk \kk'}^{\zeta \zeta'}=\frac{2\pi}{\hbar}\frac{ F(\theta) D^2}{\mathcal{A}}\frac{\hbar |\qq|^2}{2 \rho_m \omega_{\qq}} \frac{1+ \zeta \zeta' \cos \phi_{\kk \kk'} }{2} \Delta_{\kk \kk'}^{\zeta \zeta'},
\end{equation}
where the phonon momentum $\qq$ is given by the electron momentum change $\kk'-\kk$. The factor $\Delta_{\kk \kk'}^{\zeta \zeta'}$ enforces the energy conservation:
\begin{equation}
\begin{aligned}
\Delta_{\kk \kk'}^{\zeta \zeta'} &= N_{\qq} \delta(\varepsilon_{\zeta, \kk}+\hbar \omega_{\qq} -\varepsilon_{\zeta', \kk'}) \\
&+(N_{\qq}+1) \delta(\varepsilon_{\zeta, \kk}-\hbar \omega_{\qq} -\varepsilon_{\zeta', \kk'}),
\end{aligned}
\end{equation}
where $N_{\qq}$ is the bosonic phonon occupation number.

The energy-momentum conservation can be specified as:
\begin{equation}
v_F^*|\zeta' |\kk'|- \zeta |\kk||=v_{ph} |\kk'-\kk|,
\end{equation}
which requires $v_F^*>v_{ph}$ for $\zeta=\zeta'$, but $v_F^*<v_{ph}$ for $\zeta \neq \zeta'$. Therefore, acoustic phonons with velocity smaller (greater) than the Dirac Fermi velocity lead to electron intra (inter) band scattering.

Finally, the resistivity $\rho$ can be organized into the following expression:
\begin{equation}
\rho=\frac{F(\theta)D^2}{g_s g_v g_l e^2 \hbar \rho_m v_{ph}^4} \frac{k_B T}{\sum_{\zeta=\pm 1} \mathcal{M}(\zeta)}.
\label{rho_full}
\end{equation}
The dimensionless term $\mathcal{M}(\zeta)$ is the integral given by:
\begin{equation}
\mathcal{M}(\zeta)=\int_{0}^{\infty} d\varepsilon 
\frac{\varepsilon^2 f(\zeta \varepsilon)[1-f(\zeta \varepsilon)]^2}
{\int_{\varepsilon_1 '}^{\varepsilon_2 '} d \varepsilon' [1-f(\zeta' \varepsilon')] h(\zeta' \varepsilon'-\zeta \varepsilon ) \sqrt{1-x_0^2}},
\label{Mzeta}
\end{equation}
where $\zeta'=\zeta$ if $v_F^*>v_{ph}$ and $\zeta'= -\zeta$ if $v_F^*<v_{ph}$. Other terms in Eq.~(\ref{Mzeta}) are defined as follows:
\begin{equation}
\begin{aligned}
&\varepsilon_1'=\frac{|v_F^*-v_{ph}|}{v_F^*+v_{ph}}\varepsilon,\\
&\varepsilon_2'=\frac{v_F^*+v_{ph}}{|v_F^*-v_{ph}|}\varepsilon,\\
&h(E)= \frac{E|E|}{\exp[E/(k_B T)]-1},\\
&2 \varepsilon \varepsilon' x_0= \varepsilon^2+\varepsilon'^2-\Big(\frac{v_F^{*}}{v_{ph}}\Big)^{2}(\zeta' \varepsilon'-\zeta \varepsilon)^2.
\end{aligned}
\end{equation}

We calculate resistivity using Eq.~(\ref{rho_full}), and compare the results with that obtained from Eq.~(\ref{rho_TBG}). The comparison is shown in Fig.~\ref{Fig:rho_compare} for four representative twist angles. The two approaches with and without using quasielastic approximation agree quantitatively with each other for large twist angles where $v_F^*$ is at least a few times larger than $v_{ph}$ [Fig.~\ref{Fig:rho_compare}(a)]. We find that Eq.~(\ref{rho_TBG}) is accurate in predicting the slope $d\rho/d_T$ at temperature $T>T_L$ for Dirac velocity $v_F^*$ down to about $1.5 v_{ph}$ [Fig.~\ref{Fig:rho_compare}(b)]. For $v_F^* \lesssim 1.5 v_{ph}$, Equation (\ref{rho_TBG}) overestimates the resistivity  [Fig.~\ref{Fig:rho_compare}(c, d)]. We emphasize that the important qualitative features of $\rho(T) $ remain for $v_F^* \lesssim 1.5 v_{ph}$; in particular, $\rho$ is linear in $T$ for $T>T_L$ and the linear slope $d\rho/d T$ is essentially independent of the carrier density $n$, as shown in  Fig.~\ref{Fig:rho_compare}(c, d).

\section{Superconductivity theory}
\label{AppC}
We describe the detailed theoretical analysis of superconductivity in this appendix. We first elaborate on the symmetries of the system, then discuss the classification of pairing channels based on symmetries, and finally present gap equations separately for each channel.

Because we focus on small twist angle around 1$^{\circ}$, the moir\'e period $a_M$ is more than 40 times the monolayer graphene lattice constant $a_0$. As a result, the atomic scale commensurability can be neglected in the long-wavelength approximation. The moir\'e Hamiltonian in Eq.~(\ref{Hmoire}) is spatially periodic with the moir\'e period $a_M$. The moir\'e Hamiltonian also respects the $D_6$ point group symmetry. Twisted bilayer with $D_6$ symmetry is realized by  starting from AA stacked bilayer graphene, and then rotating the bottom and top layers by angles $−\theta/2$ and $+\theta/2$ around one of the hexagon centers.  Because spin-orbit coupling is negligible in graphene, there is also spin SU(2) symmetry.

In summary, the theory that we study for TBG has   
moir\'e translational symmetry, $D_6$ point group symmetry, spin SU(2) symmetry, valley U(1) symmetry and time-reversal symmetry. Because of the spin SU(2) symmetry, the pairing channels can be classified into spin singlet and spin triplet. The point group symmetries further distinguish $s$-wave from $d$-wave pairings in the singlet channels, and  $p$-wave from $f$-wave pairings in the triplet channels.

As we have discussed in Sec.~\ref{sec:SC}, acoustic phonons mediate effective attractive interactions in all four pairing channels ($s$, $p$, $d$, $f$) in TBG, mainly because of (i) sublattice pseudospin chirality and (ii) enlarged SU(2)$\times$SU(2) symmetry. Here the sublattice pseudospin chirality refers to the sublattice and valley dependent angular momenta under the three-fold rotation around $\hat{z}$ axis:
\begin{equation}
\hat{C}_3 \hat{\psi}^{\dagger}(\rr) \hat{C}_3^{-1} = \exp[i 2\pi \sigma_z \xi_z /3] \hat{\psi}^{\dagger}(\mathcal{R}_3 \rr),
\end{equation}
which can give rise to pairing channels with a finite orbital angular momentum.  

In the following, we present the linearized gap equation separately for $s$-wave, $f$-wave, $d$-wave and $p$-wave pairing channels. The gap equation faithfully incorporates the full moir\'e band structure throughout the moir\'e Brillouin zone. Because of the enlarged SU(2)$\times$SU(2) symmetry, spin singlet $s$-wave channel and spin triplet $f$-wave channel have similar gap equations, and the same is true for $d$-wave and $p$-wave channels.

{\it $s$-wave pairing.---} The $s$-wave channel has intrasublattice spin-singlet pairing. The local pair amplitude is defined as:
\begin{equation}
\Delta_{\ell}^{(s)}(\rr) = \langle \hat{\psi}_{- \sigma \ell \downarrow}(\rr)  \hat{\psi}_{+ \sigma \ell \uparrow}(\rr) \rangle = - \langle \hat{\psi}_{- \sigma \ell \uparrow}(\rr)  \hat{\psi}_{+ \sigma \ell \downarrow}(\rr) \rangle.
\end{equation}
We assume that the pair amplitude $\Delta_{\ell}^{(s)}$ is sublattice ($\sigma$) independent, but we formally keep the layer ($\ell$) dependence. We further assume that the pair amplitude has the moir\'e periodicity and therefore can be parametrized using harmonic expansion:
\begin{equation}
\Delta_{\ell}^{(s)}(\rr) = \sum_{\bb} e^{i \bb \cdot \rr} \Delta_{\bb,\ell}^{(s)},
\end{equation}
where $\bb$ is the moir\'e reciprocal lattice vectors. The linearized gap equation is then given by:
\begin{equation}
\begin{aligned}
\Delta^{(s)}_{\bb,\ell} = & \sum_{\bb'\ell'} \chi_{\bb \bb'}^{\ell \ell'} \Delta^{(s)}_{\bb',\ell'}, \\
\chi_{\bb \bb'}^{\ell \ell'}= & \frac{ 2 g_0}{\mathcal{A}} \sum_{\qq,n_1,n_2}  \Big\{  \frac{1-n_F[\varepsilon_{n_1}(\qq)]-n_F[\varepsilon_{n_2}(\qq)]}{\varepsilon_{n_1}(\qq)+\varepsilon_{n_2}(\qq)-2 \varepsilon_F} \\
& \times [\langle u_{n_1}(\qq) | u_{n_2}(\qq) \rangle_{\bb, \ell}]^* \langle u_{n_1}(\qq) | u_{n_2}(\qq) \rangle_{\bb', \ell'} \Big\},
\end{aligned}
\label{chi_s}
\end{equation}
where $\chi$ is the pair susceptibility is the $s$-wave channel,
$\qq$ labels momentum within MBZ,
$n_{1, 2}$ are moir\'e band indices in $+K$ valley for one spin component,
$\varepsilon_n$ and $|u_n\rangle$ are the corresponding energies and wave functions,
$n_F(\varepsilon)$ is the Fermi-Dirac occupation function, and $\varepsilon_F$ is the chemical potential.
The overlap function $\langle ... \rangle_{\bb, \ell}$ is the layer-resolved matrix element of the plane-wave operator $\exp(i \bb \cdot \rr )$.

We calculate $\chi$ by including momenta $\bb$ up to the third moir\'e reciprocal lattice vector shell, and by retaining only the two nearly flat bands.
The critical temperature $T_c$ is reached when the largest eigenvalue of $\chi$ is equal to 1, and the theoretical $T_c$ is shown in Fig.~\ref{Fig:Tc}.

The eigenvector of $\chi$ with the largest eigenvalue specifies the spatial and
layer dependence of the pair amplitude.  For $s$-wave pairing, the pair amplitude is layer independent
and concentrated near AA regions in the moir\'e pattern following the normal state electron density in the flat bands \cite{Wu2018phonon}.

The uniform susceptibility, i.e., the component of $\chi$ with $\bb=\bb'=0$, has the standard $g_0 \int d \varepsilon N(\varepsilon) [1-2 n_F(\varepsilon)]/[2(\varepsilon-\varepsilon_F)]$ form,
where $N(\varepsilon)$ is the DOS per spin and per valley. Therefore, the dimensionless BCS coupling constant $\lambda^*$ can be defined in the usual way as $g_0 N(\varepsilon_F)$.

{\it $f$-wave pairing.---} The $f$-wave channel has intrasublattice spin-triplet pairing. The $f$-wave channel can be further decomposed into three subchannels based on the $\hat{z}$-componet spin quantum number $S_z$:
\begin{equation}
\begin{aligned}
\Delta_{\ell}^{(f,+1)}(\rr) & = \langle \hat{\psi}_{- \sigma \ell \uparrow}(\rr)  \hat{\psi}_{+ \sigma \ell \uparrow}(\rr) \rangle, \\
\Delta_{\ell}^{(f,-1)}(\rr) & = \langle \hat{\psi}_{- \sigma \ell \downarrow}(\rr)  \hat{\psi}_{+ \sigma \ell \downarrow}(\rr) \rangle, \\
\Delta_{\ell}^{(f,0)}(\rr) & = \langle \hat{\psi}_{- \sigma \ell \downarrow}(\rr)  \hat{\psi}_{+ \sigma \ell \uparrow}(\rr) \rangle = + \langle \hat{\psi}_{- \sigma \ell \uparrow}(\rr)  \hat{\psi}_{+ \sigma \ell \downarrow}(\rr) \rangle.
\end{aligned}
\end{equation}
The above three pair amplitudes $\Delta_{\ell}^{(f,S_z)}$ with $S_z=0$ and $\pm 1$ are decoupled in the linearized gap equation, and the susceptibility for each pair amplitude $\Delta_{\ell}^{(f,S_z)}$ is the same as that for the $s$-wave channel as specified by Eq.~(\ref{chi_s}). This is due to the enlarged SU(2)$\times$SU(2) symmetry. Therefore, $s$-wave and $f$-wave channels have degenerate $T_c$ for acoustic phonon mediated pairing.

{\it $d$-wave pairing.---}
The $d$-wave channel has intersublattice spin-singlet pairing with pair amplitudes specified by:
\begin{equation}
\begin{aligned}
\Delta_{\ell}^{(+)}(\rr) &= \langle \hat{\psi}_{- B \ell \downarrow}(\rr)  \hat{\psi}_{+ A \ell \uparrow}(\rr) \rangle = - \langle \hat{\psi}_{- B \ell \uparrow}(\rr)  \hat{\psi}_{+ A \ell \downarrow}(\rr) \rangle,\\
\Delta_{\ell}^{(-)}(\rr) &= \langle \hat{\psi}_{- A \ell \downarrow}(\rr)  \hat{\psi}_{+ B \ell \uparrow}(\rr) \rangle = - \langle \hat{\psi}_{- A \ell \uparrow}(\rr)  \hat{\psi}_{+ B \ell \downarrow}(\rr) \rangle.
\end{aligned}
\label{Delta_d}
\end{equation}

We again express the pair amplitudes using harmonic expansion: $\Delta_{\ell}^{(d)}(\rr) = \sum_{\bb} e^{i \bb \cdot \rr} \Delta_{\bb,\ell}^{(d)}$, where the superscript $d=\pm$ represents the two $d$-wave pairings in Eq.~(\ref{Delta_d}). The corresponding linearized gap equation is given by:
\begin{equation}
\begin{aligned}
\Delta^{(d)}_{\bb,\ell} = & \sum_{\bb'\ell' d'} \chi^{(\bb \ell d) }_{(\bb' \ell' d')} \Delta^{(d')}_{\bb',\ell'}, \\
\chi^{(\bb \ell d) }_{(\bb' \ell' d')}= & \frac{ 4 g_0}{\mathcal{A}} \sum_{\qq,n_1,n_2}  \Big\{  \frac{1-n_F[\varepsilon_{n_1}(\qq)]-n_F[\varepsilon_{n_2}(\qq)]}{\varepsilon_{n_1}(\qq)+\varepsilon_{n_2}(\qq)-2\varepsilon_F} \\
& \times [\langle u_{n_1}(\qq) | \sigma_{d} | u_{n_2}(\qq) \rangle_{\bb, \ell}]^* \\
& \times\langle u_{n_1}(\qq) | \sigma_{d'} | u_{n_2}(\qq) \rangle_{\bb', \ell'} \Big\},
\end{aligned}
\label{chi_d}
\end{equation}
where $\sigma_{\pm}=(\sigma_x \pm i \sigma_y)/2$ are the raising and lowering operators in the sublattice space, and other terms have the same meaning as Eq.~(\ref{chi_s}).

The operators $\sigma_{\pm}$ are closely related to the velocity operator
$\partial \mathcal{H}_{+}/\partial \kk $.
Near the magic angle, the velocity of the flat bands is strongly suppressed, but
the layer counter-flow velocity remains large \cite{Bistritzer2011}.
Therefore, in the leading $d$-wave instability, the pair amplitudes are phase shifted by $\pi$ for the two layers:
$\Delta_{\mathfrak{b}}^{(d)}(\rr) = - \Delta_{\mathfrak{t}}^{(d)}(\rr)$.\cite{Wu2018phonon, wu2018topological}

We calculate the $d$-wave susceptibility in a similar way as in the $s$-wave case. The theoretical $T_c$ for $d$-wave channel is shown in Fig.~\ref{Fig:Tc}. At the critical temperature $T_c$ for $d$-wave, there are actually two degenerate  states, which can be classified into $d+i d$ and $d-i d$ chiral states. The real space texture of the pair amplitudes in the chiral states is discussed in Ref.~\cite{wu2018topological}.

{\it $p$-wave pairing.---}The $p$-wave channel has intersublattice spin-triplet pairing. Similar to $f$-wave channel, $p$-wave channel can  be decomposed into three subchannels based on the spin quantum number $S_z$. Each $p$-wave subchannel has the same pair susceptibility as  $d$-wave channel  [Eq.~(\ref{chi_d})]. Again due to the enlarged SU(2)$\times$SU(2) symmetry, $d$-wave and $p$-wave channels have degenerate $T_c$ for acoustic phonon mediated pairing.

\bibliographystyle{apsrev4-1}
\bibliography{refs}

\end{document}